\documentclass[a4paper,fleqn]{cas-sc}
\usepackage{subfigure}
\usepackage{graphicx}
\usepackage{float}
\usepackage{threeparttable}
\usepackage{hyperref}
\usepackage{url}
\usepackage{diagbox}
\usepackage{ulem}
\usepackage{caption}
\usepackage{bm}
\usepackage[authoryear]{natbib}
\usepackage{lineno} 
\def\tsc#1{\csdef{#1}{\textsc{\lowercase{#1}}\xspace}}
\tsc{WGM}
\tsc{QE}
\tsc{EP}
\tsc{PMS}
\tsc{BEC}
\tsc{DE}
\usepackage{xcolor} 
\hypersetup{
    colorlinks=true,
    citecolor={RoyalBlue},
    linkcolor={RoyalBlue},
    urlcolor={RoyalBlue}
}
\begin{document}
\begin{sloppypar}
\let\WriteBookmarks\relax
\def\floatpagepagefraction{1}
\def\textpagefraction{.001}
\shorttitle{\quad}
\shortauthors{Hao Lyu et~al.}

\title [mode = title]{Mitigating Traffic Oscillations in Mixed Traffic Flow with Scalable Deep Koopman Predictive Control}                      
\author[1,2,3]{Hao Lyu}[orcid=0000-0002-1664-8050]
\ead{lyu_hao@seu.edu.cn}

\address[1]{School of Transportation, Southeast University, Nanjing, China, 211189}
\address[2]{Jiangsu Key Laboratory of Urban ITS, Nanjing, China, 210096}
\address[3]{Jiangsu Collaborative Innovation Center of Modern Urban Traffic Technologies, Nanjing, China, 210096}
\address[4]{Department of Civil Engineering, Institute of Transport Studies, Monash University, Melbourne, Australia, VIC 3800}
\address[5]{College of Transportation Engineering, Tongji University, Shanghai, PR China, 201804}
\author[1,2,3]{Yanyong Guo}
\cormark[1]
\ead{guoyanyong@seu.edu.cn}
 
\author[1,2,3]{Pan Liu}
\ead{liupan@seu.edu.cn}

\author[4]{Nan Zheng}
\ead{nan.zheng@monash.edu}

\author[4, 5]{Ting Wang}
\ead{2110763@tongji.edu.cn}
\author[1,2,3]{Quansheng Yue}
\ead{yue_qs@seu.edu.cn}

\begin{abstract}
Mitigating traffic oscillations in mixed flows of connected automated vehicles (CAVs) and human-driven vehicles (HDVs) is critical for enhancing traffic stability. A key challenge lies in modeling the nonlinear, heterogeneous behaviors of HDVs within computationally tractable predictive control frameworks. This study proposes an adaptive deep Koopman predictive control framework (AdapKoopPC) to address this issue. The framework features a novel deep Koopman network, AdapKoopnet, which represents complex HDV car-following dynamics as a linear system in a high-dimensional space by adaptively learning from naturalistic data. This learned linear representation is then embedded into a Model Predictive Control (MPC) scheme, enabling real-time, scalable, and optimal control of CAVs. We validate our framework using the HighD dataset and extensive numerical simulations. Results demonstrate that AdapKoopnet achieves superior trajectory prediction accuracy over baseline models. Furthermore, the complete AdapKoopPC controller significantly dampens traffic oscillations with lower computational cost, exhibiting strong performance even at low CAV penetration rates. The proposed framework offers a scalable and data-driven solution for enhancing stability in realistic mixed traffic environments. The code is made publicly available
\footnote{\url{https://github.com/SpaceTrafficSafetyTeam/AdapKoopPC}}.
\end{abstract}


\maketitle

\section{Introduction}

Traffic oscillations, or stop-and-go waves, are a pervasive phenomenon in congested traffic, leading to significant detriments in traffic efficiency, fuel economy, and road safety \citep{li2014stop, chen2014periodicity}. The advent of Connected and Automated Vehicles (CAVs) offers a transformative approach to mitigate these instabilities \citep{zhao2024survey, wang2025adaptive}. By acting as mobile actuators, CAVs can be strategically controlled to actively dampen the propagation of traffic waves, thereby smoothing the overall traffic flow \citep{stern2018dissipation, liu2025optimizing, zhang2025mitigating}. However, the widespread adoption of CAVs will be a gradual process, leading to a prolonged transition period characterized by mixed traffic, where CAVs must coexist and interact with a majority of human-driven vehicles (HDVs) \citep{gan2024large, zou2025analyzing}. This complex environment presents a fundamental challenge for effective CAV control \citep{zhou2025knowledge, lai2024advancements}.

The central challenge lies in a fundamental trade-off between the high-fidelity prediction of HDV behavior and the stringent requirements of real-time, safety-critical vehicle control. On one hand, any proactive control strategy hinges on the ability to accurately anticipate the reactions of surrounding HDVs \citep{zheng2024interpretable, jiao2024digital}. Yet, human driving behavior is inherently nonlinear, stochastic, and heterogeneous, making it difficult to capture with conventional physics-based car-following models \citep{zhou2025twenty, punzo2021calibration}. While advanced data-driven models, such as those based on deep learning, have demonstrated superior accuracy in capturing these complex dynamics \citep{hart2024towards}, their non-convex and computationally intensive nature makes them ill-suited for direct integration into real-time optimization frameworks \citep{wang2023distributed, long2024traffic}. On the other hand, Model Predictive Control (MPC) stands out as an ideal framework for CAV control, as it systematically handles state and input constraints to guarantee safety and performance \citep{zhang2025preview, wang2023deep, wang2024adaptive}. The efficacy of MPC, however, is predicated on the availability of a computationally tractable—typically linear or convex—predictive model of the entire system, including the HDV responses. This creates a dilemma: high-fidelity nonlinear prediction models compromise real-time control feasibility, whereas overly simplified models fail to capture essential traffic dynamics, leading to suboptimal or unsafe CAV actions \citep{li2025robust, lyu2024kooplcc}.

To resolve this prediction-control dilemma, this paper introduces the Koopman operator theory as a powerful mathematical framework to model and predict nonlinear HDV dynamics in a manner that is amenable to real-time predictive control. The Koopman operator provides a global, linear representation of a nonlinear dynamical system by lifting the system's state into a higher-dimensional observable space \citep{zhang2024meta, lusch2018deep}. This approach allows us to decouple the complexity of modeling from the complexity of control. By leveraging deep neural networks, we can learn the intricate mapping from the original state space to this linear Koopman space, thereby capturing the nonlinear car-following behaviors of HDVs with high fidelity. The resulting model, however, is a linear system in the lifted space, which can be seamlessly and efficiently integrated into an MPC framework. This enables us to achieve both accurate, data-driven prediction and real-time, constrained optimal control.

In this paper, we propose an \textbf{Adap}tive deep \textbf{Koop}man \textbf{P}redictive \textbf{C}ontrol framework (AdapKoopPC) for regulating mixed traffic flow. This framework is designed to learn heterogeneous HDV behaviors from naturalistic driving data and deploy scalable, real-time control strategies for CAVs to mitigate traffic oscillations. The main contributions of this work are threefold:

\begin{itemize}
    \item We develop a deep Koopman network-based trajectory prediction model, termed \textbf{AdapKoopnet}, which adaptively identifies latent driving scenarios and extracts personalized driver characteristics from unlabeled, naturalistic trajectory data to deliver highly accurate HDV behavior predictions.
    \item We propose the scalable \textbf{AdapKoopPC} control framework, which embeds the linear predictive blocks from pre-trained AdapKoopnet models into an MPC formulation. This framework can be dynamically constructed based on the local vehicle composition, enabling real-time, cooperative control of CAVs in arbitrary mixed traffic configurations.
    \item We conduct extensive simulations to validate our framework's performance. The results confirm that AdapKoopPC markedly reduces traffic oscillations and enhances traffic stability in both small- and large-scale systems (10 and 50 vehicles, respectively), even at low CAV penetration rates, while maintaining computational feasibility for real-time deployment.
\end{itemize}

The remainder of this paper is organized as follows. Section 2 reviews related literature and highlights the research gap. Section 3 details the architecture of the AdapKoopnet for HDV trajectory prediction. Section 4 presents the AdapKoopPC framework for mixed traffic control. Sections 5 and 6 present experimental validation for the prediction and control models, respectively. Finally, Section 7 concludes the paper and discusses future research directions.

\section{Related work}

\subsection{Car-Following Behavior Modeling and Prediction of HDVs}
Accurate prediction of HDVs behavior is a critical prerequisite for the effective control of CAVs in mixed traffic. The challenge is to develop a model that is not only descriptively powerful but also computationally compatible with real-time, optimization-based control frameworks.
Historically, HDV modeling has relied on physics-based car-following models, such as the Intelligent Driver Model (IDM) and the Optimal Velocity Model (OVM) \citep{treiber2000congested, bando1998analysis, kim2024asymmetric}. While these models offer an interpretable structure, their fixed parameters struggle to capture the profound nonlinearity and heterogeneity of real-world driving behavior, often leading to significant prediction errors \citep{zhang2024car, zhang2024calibrating}. To address this accuracy deficit, researchers have increasingly adopted data-driven methods. Deep learning models, particularly Long Short-Term Memory (LSTM) networks, excel at learning complex temporal dependencies from trajectory data and have achieved state-of-the-art prediction accuracy \citep{xu2024sequence, zhang2019simultaneous}. However, their "black-box" nature and computationally intensive nonlinear structures are fundamentally ill-suited for direct integration into real-time control frameworks, which demand tractable and often linear models for rapid optimization.
To further enhance model fidelity, another line of research focuses on personalizing predictions by incorporating human factors. These efforts include identifying distinct driving styles, such as aggressiveness \citep{zhang2022generative, chen2024aggfollower}, or clustering behaviors into different traffic scenarios \citep{song2023personalized, li2021extraction}. While such methods improve predictive power by accounting for driver- or situation-specific nuances, they often introduce additional complexity without resolving the underlying conflict between model nonlinearity and real-time control feasibility. Some hybrid approaches, like Physics-Informed Neural Networks (PINNs), attempt to find a balance by embedding physical laws into the learning process but still typically yield nonlinear representations \citep{mo2021physics, kang2023trajectory}.

Therefore, a critical gap remains in HDV behavior modeling: the need for a framework that can learn complex, personalized driving behaviors from data, yet yields a predictive model with a computationally tractable structure suitable for real-time applications. Existing paradigms force a trade-off between predictive fidelity and computational feasibility, motivating the search for novel modeling approaches that can bridge this divide.

\subsection{CAV Control Strategies for Mitigating Traffic Oscillations}

Building upon the challenge of accurately modeling HDV behavior, the next critical step is to design effective control strategies for CAVs that can actively mitigate traffic oscillations. The paradigm for such controllers has evolved from localized, reactive maneuvers to holistic, predictive frameworks that optimize the behavior of the entire mixed-traffic system \cite{wang2023distributed, wang2023general}. Early strategies sought to create spatial buffers to absorb traffic waves. These range from pre-planned maneuvers like Jam-Absorption Driving (JAD), where a vehicle is guided to slow down in anticipation of a jam \citep{nishi2013theory, he2016jam}, to specific control laws like the Follower Stopper (FS) controller, which uses larger-than-normal headways to smooth traffic and has been validated in field experiments \citep{stern2018dissipation, wang2024mitigating}. While conceptually insightful, such methods are often limited in their adaptability, motivating a shift towards more proactive and comprehensive control frameworks like Leading Cruise Control (LCC) \citep{wang2021leading}. In the LCC framework, a CAV acts as both a follower adapting to preceding vehicles and a leader actively influencing following vehicles, empowering it to exert system-level control to stabilize traffic flow \citep{zheng2020smoothing, shang2024decentralized}.

The implementation of the LCC concept has been explored through three principal control methodologies, each presenting a distinct trade-off between model fidelity, computational tractability, and safety assurance. (1) \textbf{MPC} is a natural fit for LCC due to its inherent ability to handle complex system dynamics and enforce hard constraints on states and inputs, ensuring safety and comfort \citep{zhang2025mixed}. Its performance, however, is fundamentally tethered to the accuracy of the underlying predictive model, which typically requires an explicit, often linearized, model of HDV behavior \citep{wang2021leading}. This inevitably creates a mismatch with the true nonlinear dynamics of human drivers, potentially degrading control performance and robustness. (2) \textbf{Data-Driven Predictive Control} aims to circumvent this reliance on explicit models. A prominent example, the Data-EnablEd Predictive Leading Cruise Control (DeeP-LCC) framework, constructs a non-parametric model directly from historical trajectory data \citep{wang2023deep}. While effective, this approach requires large amounts of high-quality offline data that must be representative of online conditions, and the resulting optimization can be computationally demanding for large-scale systems \citep{wang2023distributed, shang2024decentralized}. (3) \textbf{Reinforcement Learning (RL)} offers a powerful model-free alternative, capable of learning complex, nonlinear control policies through interaction with the environment \citep{li2024augmented, xu2024multi, shi2025predictive}. Despite its adaptability, the principal drawback of RL is the profound difficulty in providing formal safety guarantees. Encouraging safety through reward shaping does not offer assurance against violating hard state constraints, a critical limitation for safety-critical applications like autonomous driving \citep{zhou2024enhancing}.

In summary, the implementation of advanced CAV control faces a methodological trilemma: 1) MPC offers robust constraint handling but is hindered by model mismatch with real-world HDV behavior; 2) Data-driven methods like DeeP-LCC avoid explicit modeling but are data-intensive and computationally demanding; and 3) Reinforcement Learning excels at learning complex policies but struggles to provide formal safety guarantees. This highlights the need for a holistic control framework that synergizes the strengths of these approaches—namely, the data-driven adaptability, the rigorous constraint satisfaction of MPC, and real-time computational efficiency. Our development of a Koopman operator-based predictive control framework is directly motivated by this systemic need.

\subsection{The applications of Koopman operator theory in traffic flow} 
The Koopman operator theory emerges as a powerful framework to reconcile the aforementioned trade-offs by representing nonlinear dynamics with a linear operator in a higher-dimensional space \citep{wang2024koopman, lusch2018deep}. Its key advantage lies in constructing an explicit, data-driven linear predictive model from complex trajectory data, thereby bridging the gap between high-fidelity nonlinear models and the computational demands of real-time MPC \citep{xiao2022deep, chen2024data}. Capitalizing on these advantages, researchers have begun applying Koopman operator theory across various scales of traffic control. At the macroscopic level, it has been used to analyze and forecast complex network-wide traffic dynamics \citep{avila2020data} and to design efficient controllers for signalized intersections and freeway ramp metering \citep{ling2020koopman, gu2023deep, das2023koopman}. More pertinent to this work, its application has extended to the microscopic level for modeling and controlling vehicle platoons. Studies have successfully employed Koopman-based methods to create data-driven models of mixed-vehicle platoons, combining deep learning with physical interpretability to enhance accuracy and stability analysis \citep{tian2024physically, zhan2022data}. Crucially, recent work has leveraged these learned linear models as the predictive core within MPC frameworks to achieve robust, real-time control of CAVs in mixed traffic \citep{li2025robust, 10878999}.

Despite this progress, existing Koopman-based control strategies exhibit two key limitations that hinder their practical scalability. First, their validation often relies on trajectory data generated from idealized car-following simulations, which may not capture the full complexity of real-world driving behaviors. Second, and more critically, current methods typically learn a single, monolithic Koopman operator for a fixed platoon configuration (i.e., a specific number and arrangement of vehicles). This static approach lacks adaptability; if the traffic composition changes, the model becomes invalid. This necessity for a scalable and adaptive framework, capable of learning from naturalistic data and handling dynamic traffic formations, directly motivates the contributions of this paper.

\section{AdapKoopnet: Adaptive deep Koopman network for car following behavior modeling and prediction of HDVs}
In this section, a data-driven adaptive deep Koopman linear model is proposed to address the challenges associated with real-time cognition and prediction of the state of HDVs.

\subsection{Key terms definition}
Considering that terms such as scenarios have different understandings in existing research. Here we define and explain several key terms that apply specifically to this paper: 

\textit{Scenario} is the mixed traffic flow environment in which the vehicle is located, such as free flow, synchronous flow, congestion flow, etc., and the direct explicit status includes traffic flow velocity, density, etc. 

\textit{Scenario characteristics} refers to the collective driving behavior exhibited by drivers in corresponding scenes, which is a potential common feature. For example, in high-velocity and high-density driving scenario, drivers generally pay more attention to the behavior of surrounding vehicles, are greatly influenced by them, and adjust their own driving behavior more frequently than usual.

\textit{Driving characteristics} is the specific manifestations of an individual driver's long-term driving habits in different driving scenarios. For example, aggressive drivers may be more conservative in high-velocity and high-density scenarios compared to free flow scenarios. However, this tendency is uncertain. Some drivers exhibit driving characteristics similar to the average of vehicle group characteristics in certain scenarios, while others are only slightly affected. Therefore, the predicted scenario classification is the comprehensive value of driver tendency and scenario characteristics themselves.

\subsection{Problem description}
Assuming that in high-density mixed traffic flow, HDVs not engaging in lane-changing behavior are primarily influenced by their preceding vehicles. Their driving behavior is primarily shaped by the current velocity of the preceding vehicle, their own current velocity, and the headway. For the purpose of research, time is discretized into infinitesimally small segments, and the aforementioned process can be described by the following equations\citep{saifuzzaman2014incorporating}: 
\begin{equation}
\left( {{v_i}\left( {t + 1} \right),{h_i}\left( {t + 1} \right)} \right) = f\left( {{v_i}\left( t \right),{h_i}\left( t \right),{v_{i - 1}}\left( t \right)} \right)
\end{equation}
where ${v_i}\left( t \right),{h_i}\left( t \right),{v_{i - 1}}\left( t \right)$ respectively represent the velocity, headway of vehicle $i$, and the velocity of the preceding vehicle $i-1$ at time $t$; $f\left( {} \right)$ denotes the state transition function.

In real-world scenarios, different drivers commonly exhibit markedly diverse behaviors when faced with identical situations. This variability is intricately associated with individual driving habits, short-term fluctuations in the surrounding environment, and specific driving intentions.

These short-term trajectories serve as external manifestations of driver characteristics, encapsulating abundant driving semantic information. Consequently, they are employed for the identification and differentiation of heterogeneity among drivers. Therefore, the problem is defined as follows:
\begin{equation}
\begin{array}{l}
\label{eq2}
{dc_i}\left( t \right) = {f_{dc}}\left( {{x_i}\left( {t - P} \right),{x_i}\left( {t - P + 1} \right),{x_i}\left( t \right)} \right)\\
\left( {v_i}\left( {t + 1} \right),{h_i}\left( {t + 1} \right)\right) = {f_{sp}}\left( {{v_i}\left( t \right),{h_i}\left( t \right),d{c_i}\left( t \right),{v_{i - 1}}\left( t \right)} \right)
\end{array}
\end{equation}
where ${x_i}\left( \cdot \right) = \left[ {{v_i}\left( \cdot \right),{h_i}\left(  \cdot\right),\Delta {v_i}\left( \cdot \right),{a_i}\left( \cdot \right),{l_i}} \right]$ , $\Delta {v_i}\left( \cdot \right),\;{a_i}\left( \cdot \right),\;{l_i}$  respectively represent the velocity difference, acceleration, and vehicle length of vehicle ${i}$; $P$, ${f_{dc}}\left( \cdot \right)$ ,$d{c_i}\left(  \cdot  \right)$   respectively represent the length of historical trajectories, the mapping relationship between trajectories and driving characteristics, and the driving characteristics extracted from information containing $P$ trajectory samples.

As illustrated in Eq. (2), the objective in this section revolves  finding a mapping. The inputs contain the historical trajectory context, the current explicit state of the vehicle $i$, the velocity of the preceding vehicle $i-1$, and the outputs contain the prediction velocity and headway of the vehicle $i$ in next time step. However, the mapping is typically nonlinear, leading to significant computational delays in online optimization for mixed traffic flow. The Koopman operator theory provides an promising approach to tackle this challenge.

\subsection{Koopman operator theory for state prediction of HDVs}
\subsubsection{Koopman operator theory}
\begin{figure}[pos=htbp]
    \vspace{1em}  
    \centering
    \includegraphics[width=1\textwidth]{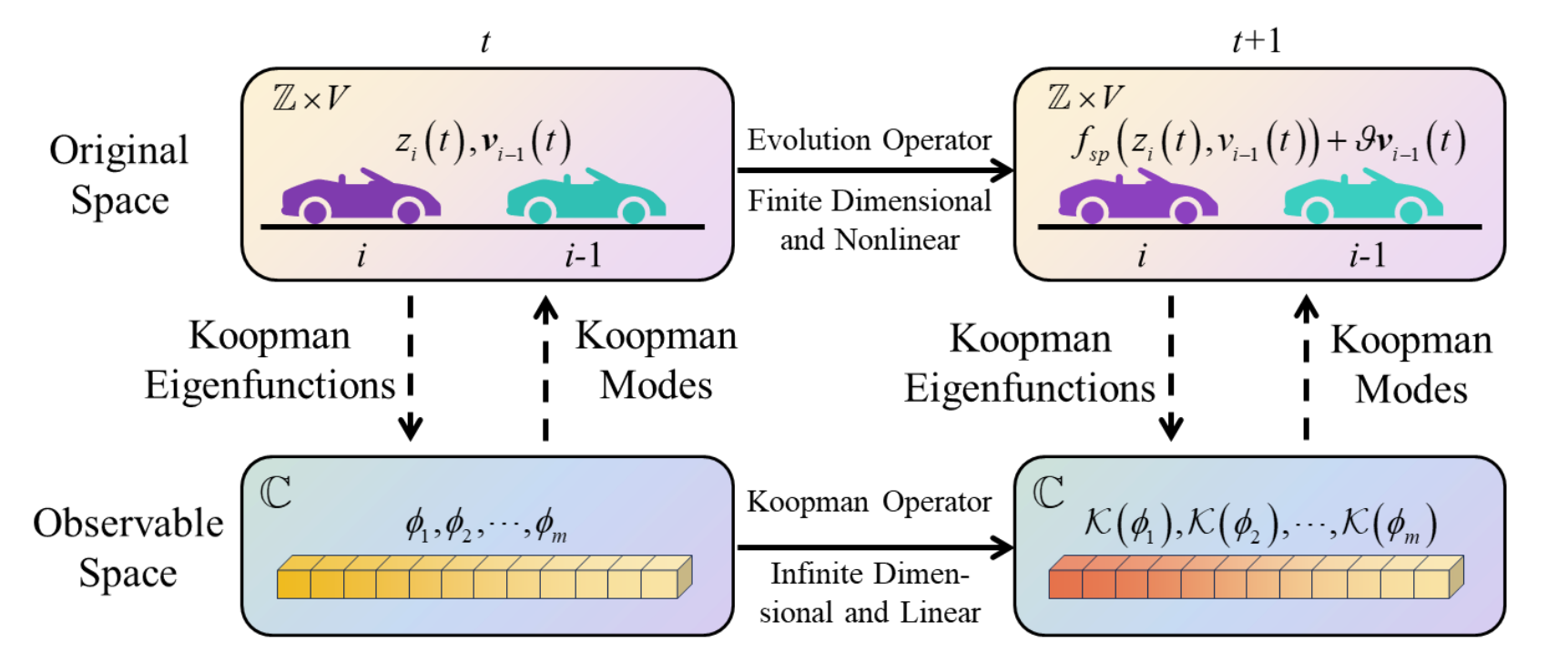}
    \vspace{-1em}
    \caption{\centering{ The relationship between the original space and the observable space }}
	\label{Figure 1}
    \vspace{-1em}  
\end{figure}

The Koopman operator theory initially provides an alternative linear dynamic description for the evolution of uncontrollable systems \citep{koopman1931hamiltonian}.  With slight modifications, the Koopman operator can be applied to controlled systems \citep{proctor2018generalizing}.   Therefore, the evolution of system modeled by Eq. (2) can be expressed by a linear Koopman operator in an infinite-dimensional space. Let ${z_i}\left(  \cdot  \right) = {\left[ {{v_i}\left(  \cdot  \right),{h_i}\left(  \cdot  \right),d{c_i}\left(  \cdot  \right)} \right]^T} \in \mathbb{Z}$   represents the state of vehicle ${i}$, ${\bm{v}_{i - 1}}\left( t \right) = v_0^\infty$ denotes all the velocities in the velocity space $V$, the Koopman operator on System corresponding to Eq.(2) with the extended state $\left[ {{z_i}\left( t \right),{\bm{v}_{i - 1}}\left( t \right)} \right]$  is defined as follows:
\begin{equation}
{{\cal K}}\phi \left( {{z_i}\left( t \right),{\bm{v}_{i - 1}}\left( t \right)} \right) = \phi \left( {{z_i}\left( {t + 1} \right),{\bm{v}_{i - 1}}\left( {t + 1} \right)} \right) = \phi \left( {{f_{sp}}\left( {{z_i}\left( t \right),{v_{i - 1}}\left( t \right)} \right) + \vartheta {\bm{v}_{i - 1}}\left( t \right)} \right)
\end{equation}
where ${\cal K}$ is the Koopman operator in the infinite-dimensional space;  $\vartheta {\bm{v}_{i - 1}}\left( t \right) = {\bm{v}_{i - 1}}\left( {t + 1} \right)$ with $\vartheta $ being a left shift operator. It is noteworthy that, unlike ${f_{dc}}\left(  \cdot  \right)$  directly acting on the state ${z_i}\left(  \cdot  \right)$, the Koopman operator ${{\cal K}}$ operates on the state space functions $\phi \left( \cdot  \right)\in \mathbb{Z}\times V$  with $\phi $ : $\mathbb{Z}\times V\to \mathbb{C}$ . Exploiting the linearity of ${{\cal K}}$, it can be subjected to eigenvalue decomposition, expressed as follows:
\begin{equation}
{{\cal K}}{\phi _m}\left( {{z_i}\left( t \right),{{v}_{i - 1}}\left( t \right)} \right) = {\lambda _m}{\phi _m}\left( {{z_i}\left( t \right),{{v}_{i - 1}}\left( t \right)} \right)
\end{equation}
where ${\lambda _m},\;{\phi _m}\left(  \cdot  \right)$ represent the eigenvalues of ${{\cal K}}$ and their corresponding eigenfunctions, respectively. The future states of the system can be acquired either by directly evolving ${z_i}\left(  \cdot  \right)$ or by evolving the complete observable state through the Koopman operator:
\begin{equation}
{f_{sp}}\left( {{z_i}\left( t \right),{{\bm{v}}_{i - 1}}\left( t \right)} \right) = \sum\limits_{m = 1}^\infty  {{\lambda _m}{\nu _m}{\phi _m}\left( {{z_i}\left( t \right),{{\bm{v}}_{i - 1}}\left( t \right)} \right)} 
\end{equation}
where ${\nu _m}$ is the Koopman mode corresponding to the eigenvalue  ${\lambda }_{m}$. 
The relationship between the original space and the observable infinite-dimensional space is depicted in Fig. \ref{Figure 1}.

Based on Eqs. (3)-(5), the endeavor to derive a global linearized dynamic description equivalent to system modeled by Eq.(2) involves the search for Koopman eigenvalues along with their corresponding eigenfunctions and Koopman modes. Nevertheless, the Koopman operator typically encompasses an infinite number of eigenvalues. Consequently, in most instances, a global linear approximation of the system can only be achieved by identifying essential eigenvalues and their associated eigenfunctions and Koopman modes.

\subsubsection{Extended dynamic mode decomposition} 
The extended dynamic mode decomposition (EDMD) is a data-driven approach of finding the finite-dimensional approximation $\mathit{K}$ of the Koopman operator \citep{williams2015data}. EDMD employs various basis functions, such as Radial Basis Functions (RBF) with different kernel centers and widths, to represent observable functions. It utilizes least squares regression to calculate $\mathit{K}$. For forced dynamics model in Eq. (2), a special way for selecting basis functions is defined in Eq. (5) to obtain :
\begin{equation}
\phi \left( {{z_i}\left( t \right),\bm{{v}_{i - 1}}\left( t \right)} \right) = {\left[ {\varphi {{\left( {{z_i}\left( t \right)} \right)}^T}{\rm{ }},\bm{{v}_{i - 1}}{{\left( t \right)}^T}} \right]^T}
\end{equation}
where $\varphi \left( \cdot  \right)={{\left[ {{\varphi }_{1}}{{\left( \cdot  \right)}^{T}}\text{ }{{\varphi }_{2}}{{\left( \cdot  \right)}^{T}}\text{ }\cdots \text{ }{{\varphi }_{L}}{{\left( \cdot  \right)}^{T}} \right]}^{T}}$ represents a set of observable lift functions is general nonlinear. Let ${{s}_{i}}\left( k \right)=\varphi \left( {{z}_{i}}\left( k \right) \right)$, combining Eqs. (3)-(6), Eq. (7) is obtained:
\begin{equation}
\phi \left( {{z_i}\left( {k + 1} \right)} \right) = {{\cal K}}\phi \left( {{z_i}\left( k \right),{v_{n - 1}}\left( k \right)} \right) = {K}{\left[ {{s_n}{{\left( k \right)}^T}{\rm{ }}{v_{n - 1}}{{\left( k \right)}^T}} \right]^T} + r
\end{equation}
where $r$ is residual term that describes the gap between the \textit{L}-dimensional approximation of the observable space and the actual lifted space of the Koopman operator, used to determine the optimal $\mathit{K}$. However, selecting the lifting functions for the complex dynamics of System (\ref{eq2}) poses a challenge, and advanced deep learning techniques are employed to learn $\mathit{K}$ .

\subsection{Model architecture}
A deep learning model, based on attention mechanisms and feedforward networks, is constructed to accomplish the following tasks: 1) Extracting latent driving characteristics from historical trajectory context, which are utilized to aid in understanding and predicting the behavior of HDVs; 2) Learning Koopman lifting functions, Koopman operator approximation, and Koopman modes, the latter two of which are linear, for online optimization in CAVs .

\begin{figure}[pos=htbp]
    \vspace{1em}  
    \centering
    \includegraphics[width=1\textwidth]{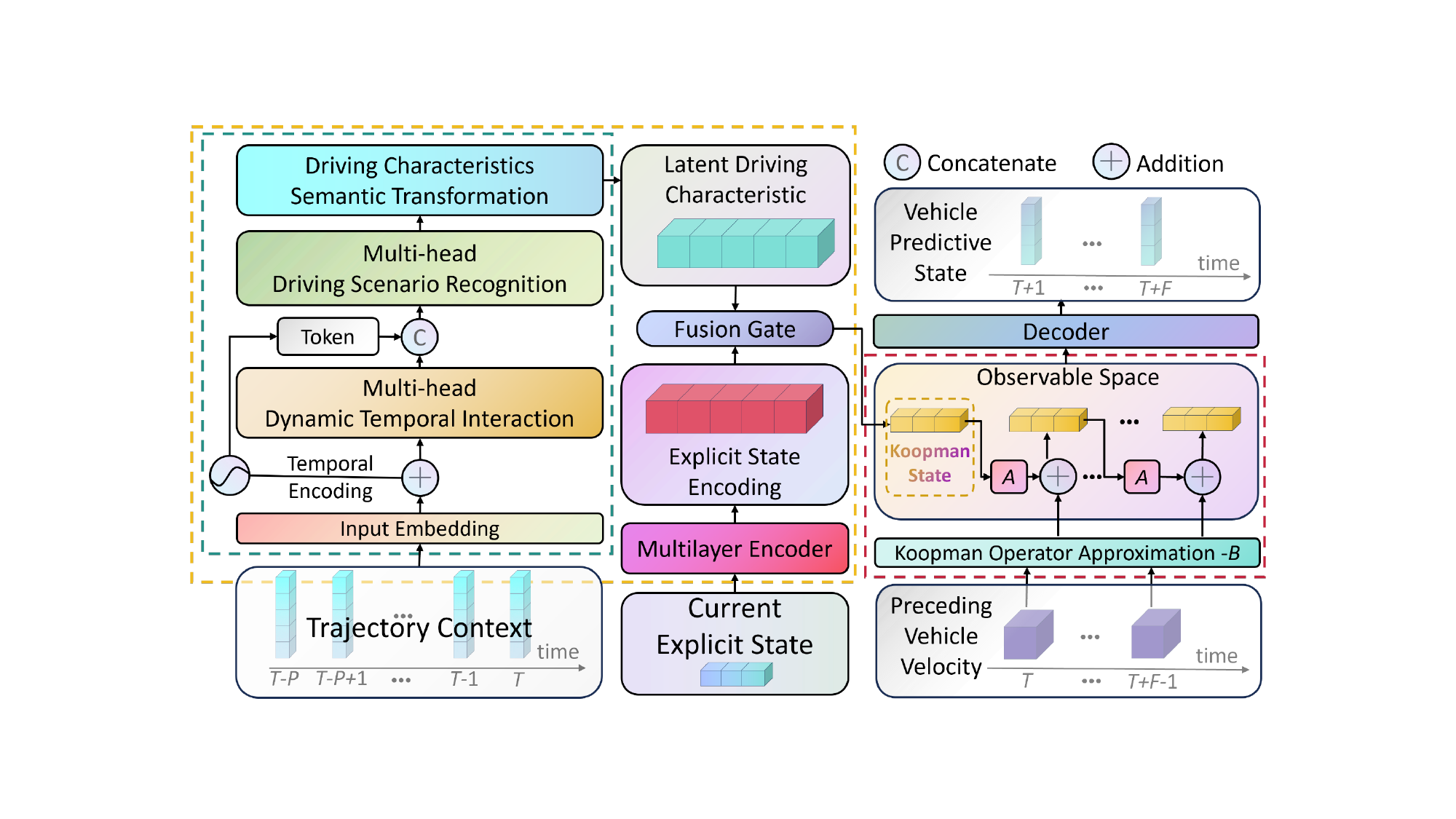}
    \vspace{-1em}
    \caption{\centering{ The model architecture of AdapKoopnet}}
	\label{Figure 2}
    \vspace{-1em}  
\end{figure}

The model architecture is depicted in Fig.\ref{Figure 2}. For trajectory context inputs, the model incorporates a driving characteristic semantic extraction block (highlighted by the deep green dashed box in Fig.\ref{Figure 2}). To handle the current vehicle state input, a multi-layer perceptron-based encoder-lifting function approximation is employed. The fusion gate mechanism integrates the encoding of driving characteristics with the explicit state encoding of the lifting space, yielding the state of the observable high-dimensional Koopman space approximation. By incorporating the future velocity of the preceding vehicle into the network, the model adaptively learns the Koopman operator approximation and achieves multi-step state predictions in the high-dimensional space. This module is denoted as the Koopman evolution block, highlighted by the pink dashed box in Fig.\ref{Figure 2}). Additionally, a linear decoder, serving as an approximation for Koopman modes, is utilized to transform observations from the high-dimensional space to obtain predicted states in the original space.

\subsubsection{Driving characteristic semantic extraction block}
As shown in Fig. \ref{Figure 2}, this block takes trajectory context as input, containing details as specified in the Eq. (2), and outputs a vector representing the current latent driving characteristics of the driver. The block comprises an input embedding and temporal encoding (ITE) module, a multi-head dynamic temporal interaction (DTI) module, a multi-head driving scenario recognition (DSR) module, and a driving characteristics semantic transformation (DCSE) module, each of which is detailed below.

\vspace{1em}
\textbf{Input Embedding and Temporal Encoding (ITE) Module}
The input embedding layer serves to convert the trajectory context into a high-dimensional dense representation, allowing the model to comprehensively learn the trajectory features and analyze correlation between trajectory samples. The embedding is achieved through a fully connected layer with a ReLU activation function. Given that the embedding operation is conducted for each trajectory sample, a relative time encoding method is introduced to enable subsequent modules to recognize the temporal information of trajectory context \citep{vaswani2017attention}. The expression is as follows:

\begin{equation}
\begin{matrix}
  T{{E}_{(t,2i)}}=\sin \left( \frac{t}{{10000}^{(2i/{{d}_{\text{model}}})}} \right) \\ 
  T{{E}_{(t,2i+1)}}=\cos \left( \frac{t}{{10000}^{(2i/{{d}_{\text{model}}})}} \right) \\ 
\end{matrix}
\end{equation}
where $i$, $t$ represent the encoding feature dimension index and time step, respectively;   ${{d}_{\text{model}}}$ represents the dimensionality of the encoding of each module in the AdaptKoopnet, without special declaration. Subsequently, the trajectory context input embedding and temporal encoding are added together to form the output of this module. The expression is as follows:
\begin{equation}
{{H}_{ITE}}=\operatorname{Re}\text{LU}\left( {{W}_{IE}}\cdot \left( x\left( T-P \right),x\left( T-P+1 \right),x\left( T \right) \right)+{{b}_{IE}} \right)+TE
\end{equation}
where ${{W}_{IE}}$, ${{b}_{IE}}$, represent the weights and biases of the input embedding layer.

\vspace{1em}
\textbf{Multi-Head Dynamic Temporal Interaction (DTI) Module}
The vehicle state undergoes continuous changes, and there exists a strong interaction and correlation between trajectory samples. This module leverages multi-head attention and feedforward layers to capture and understand the temporal interactions and dependencies within the vehicle trajectory context.
Firstly, the multi-head attention mechanism projects the output of ITE into multiple subspaces. In each subspace, it independently learns interaction features within the trajectory context and facilitates feature exchange. By focusing on different subspaces, the model can better capture information from different dimensions within the trajectory context, enhancing its ability to model complex relationships between trajectory sequences. This approach not only accelerates the speed of training and inference but also contributes to a more comprehensive understanding of the intricate dynamics among trajectory samples.

The dimension of each subspace is also a hyperparameter. In AdapKoopnet, a uniform subspace dimension is adopted and denoted as ${d_{att}}$. The calculation formula for $\bar{H}_{DTI}^{{}}$  in the figure is as follows:
\begin{equation}
\begin{array}{l}
Q_{DTI}^h = W_{DTI - q}^hH_{ITE}^{},\quad Q_{DTI}^h = W_{DTI - k}^hH_{ITE}^{},\quad Q_{DTI}^h = W_{DTI - v}^hH_{ITE}^{}\\
\bar H_{DTI}^h = {\rm{softmax}}\left( {\frac{{Q_{DTI}^hK{{_{DTI}^h}^T}}}{{\sqrt {{d_{att}}} }}} \right)V_{DTI}^h\\
\bar H_{DTI}^{} = {{LN}}\left( {{W_{DTI - att}}\left( {\left\| {_{g = 1}^H\bar H_{DTI}^h} \right.} \right) + H_{ITE}^{}} \right)
\end{array}
\end{equation}
where $W_{DTI-q}^{h}$, $W_{DTI-k}^{h}$, $W_{DTI-v}^{h}$, ${{W}_{DTI-att}}$ represents learnable weights; $\left\|\cdot\right.$ denotes the concatenate operation. $LN\left( \cdot  \right)$ stands for Layer Normalization, a technique that normalizes the trajectory encoding along the feature dimension to mitigate the impact of internal covariate shift \citep{ba2016layer}. Unlike Batch Normalization (BN), Layer Normalization is more flexible as it is not influenced by the size of the data batch.

After achieving feature exchange within the trajectory context through the attention layer, the feedforward layer \citep{fine2006feedforward} is employed for the nonlinear transformation of trajectory encoding. This aims to capture the nonlinear relationships within the trajectory context, facilitating the model in learning higher-level abstract representations. The feedforward layer consists of two linear transformations and an activation function. Initially, $\bar{H}_{DTI}^{{}}$ undergoes a fully connected layer, followed by the application of the ReLU activation function, and finally passes through another fully connected layer. After $\bar{H}_{DTI}^{{}}$ goes through the feedforward layer, the output of the DTI module is obtained, calculated using the following formula:
\begin{equation}
{H_{DTI}} = {{LN}}\left( {W_{DTI - FFN}^2\left( {{\mathop{\rm Re}\nolimits} {\rm{LU}}\left( {W_{DTI - FFN}^1\left( {{{\bar H}_{DTI}}} \right)} \right)} \right) + {{\bar H}_{DTI}}} \right)
\end{equation}
where $W_{DTI-FFN}^{1}$ and $W_{DTI-FFN}^{2}$ represent the weights of the feedforward layer in the DIT module.

\vspace{1em}
\textbf{Multi-Head Driving Scenario Recognition (DSR) Module}
Following the extracting and initial abstracting of temporal interaction characteristics within the trajectory context through the DTI module, the cognitive understanding of driving scenario and the extraction of driving characteristic semantics become crucial steps. This is because, in different driving scenarios, even for the same driver, driving characteristics may vary. For example, in scenarios with large headway, drivers may not require to remain highly vigilant about their preceding vehicles, and their driving behavior tends to be smoother. Conversely, in high-density and high-velocity scenarios, drivers may concentrate more on monitoring changes in the state of preceding vehicle and respond more actively. The DSR module learns scenario information hidden within the trajectory context by adapting to relevant feature variations from a vast set of driving trajectory contexts. It dynamically recognizes and classifies the driving scenario in which the vehicle is situated.

\begin{figure}[pos=htbp]
    \vspace{1em}  
    \centering
    \includegraphics[width=1\textwidth]{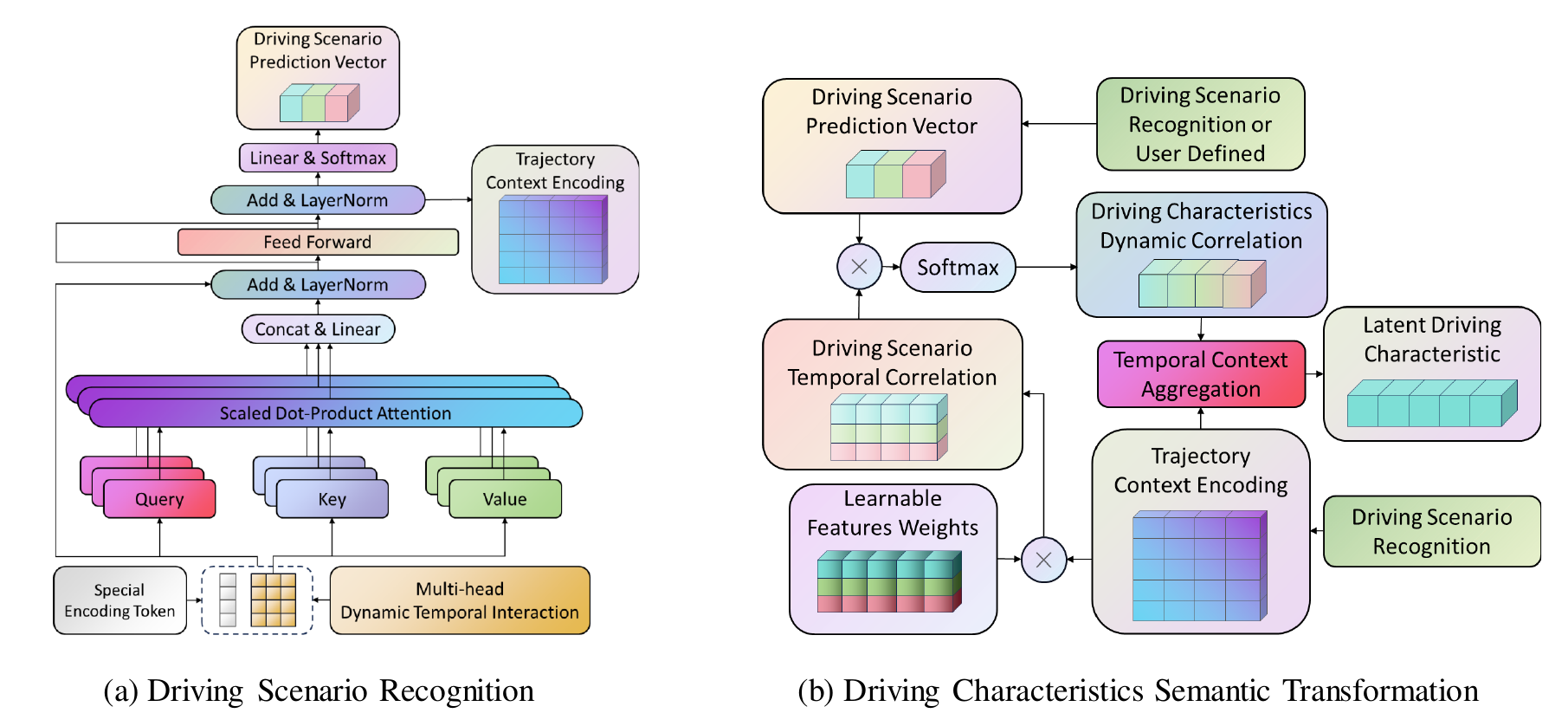}
    \vspace{-1em}
    \caption{\centering{ The architecture of DSR module and DCSE module}}
	\label{Figure 3}
    \vspace{-1em}  
\end{figure}

Fig. \ref{Figure 3} (a) illustrates the structure of the DSR module. Initially, Eq. (8) is utilized to generate a special encoding token. Subsequently, this token is concatenated with the output of the DTI module  $H_{DTI}^{{}}$, functioning as the original query. The role of this token is to extract features conducive to the cognitive understanding of driving scenarios by attending to the trajectory context encoding. These features undergo further abstraction through the nonlinear transformation of the feedforward layer. The module was associated and matched these abstracted features with the learned scenario information features of the model, achieving cognitive recognition and prediction of implicit driving scenarios within the trajectory context. Similar to the DTI module, this module conduct deep exchange and further abstraction of interaction features within the trajectory context.

The computational process of the DSR module is fundamentally similar to the DTI module. Its output in the feedforward layer is as follows:
\begin{equation}
{H_{DSR - DS}},H_{DSR - TC}^{} = {\bar W_{DSR}}\left( {SE,H_{DTI}^{}} \right)
\end{equation}
where $SE$, ${{\bar{W}}_{DSR}}$, $H_{DSR-TC}^{{}}$ and ${{H}_{DSR-DS}}$ represent the special encoding token, the weights of DSR module, the trajectory context encoding in Figure 4, and the special encoding token that completes the extraction of driving scenario features, respectively. ${{H}_{DSR-DS}}$ undergoes a linear layer to aggregate driving scenario features, resulting in a vector with the same dimension as the predefined number of scenes. After applying the softmax activation function, the driving scenario prediction vector depicted in Fig. \ref{Figure 3} (a) is obtained. Each dimension of this vector represents the predicted probability of the trajectory context belonging to the corresponding scene. The formula is expressed as follows:
\begin{equation}
{{H}_{DS}}=\text{softmax}\left( {{W}_{DSR-DS}}\cdot \left( {{H}_{DSR-DS}} \right) \right)
\end{equation}

\textbf{Remark 1:} In this study, real labels for driving scenario are not available. Therefore, the driving scenario recognition process involves a spontaneous classification process by the neural network based on a large amount of trajectory context. At the inception of the network design, the task of the DSR module is to cluster recurring trajectory patterns and interpret the generated clusters as potential scenarios. From a macro perspective, this lays the foundation for extracting semantic characteristics of driving characteristics. In Section 5, explicit features corresponding to each scenario will be visualized, although this may not necessarily be the sole basis for the classification of network.

\vspace{1em}
\textbf{Driving Characteristics Semantic Transformation (DCSE) Module}
As mentioned in the DSR module, driving characteristics vary across different driving scenarios. The current module aims to perform driving characteristic semantic extraction guided by the predicted results of driving scenario recognition. The structural diagram of module is presented in Fig. \ref{Figure 3} (b) . 

The process begins by defining learnable feature weights corresponding to each scenario. These weights dynamically learn the importance of each feature in the trajectory context encoding under predefined scenarios, using an extensive training data. The features are then aggregated based on the learned importance, obtaining the relevance of each trajectory sample to driving characteristic extraction in a specific driving scenario. Then, the relevance for each scenario is aggregated based on the probabilities predicted by the DSR module or according to user-defined scenarios. After applying the softmax activation function, the correlation between the trajectory context encoding and driving features is obtained. Based on this correlation, the trajectory context is aggregated, ultimately revealing the driving features hidden within the trajectory context. The computational formula for this module is as follows:
\begin{equation}
dc\left( T \right) = {\rm{softmax}}{\left( {H_{DSR - TC}^{}d{s_{fc}}{H_{DS}}} \right)^T}H_{DSR - TC}^{}
\end{equation}
where $d{{s}_{fc}}$, $d{{c}_{T}}$ respectively represent the learnable feature weights corresponding to each scenario, and the latent driving characteristic at time $T$.
\vspace{1em}
\subsubsection{Explicit state encoder and fusion gate mechanism}
The explicit state encoder is tasked with encoding the explicit state of the vehicle from the original space to an observable high-dimensional space, facilitating the action of the Koopman lifting function on the current vehicle state. This encoder initially embeds the vehicle state into a high-dimensional representation through a linear layer with a ReLU activation function. Subsequently, a Multi-Layer Perceptron (MLP) with tanh activation function is utilized to perform multiple linear transformations on the embedded representation, completing the task of encoding the explicit state of the vehicle. Next, the explicit state encoding and the latent driving characteristics of the vehicle need to be fused to form the Koopman state approximation for the current time step. The Gated Linear Unit (GLU) is introduced to accomplish this fusion task. It employs a gate mechanism to adaptively filter redundant features and retain essential features \citep{dauphin2017language}.
The computational formula for the above process is as follows:
\begin{equation}
\begin{array}{l}
es\left( T \right) = {W_{ESE}} \cdot es\left( T \right)\\
s\left( T \right) = {W_{FG}} \cdot \left( {dc\left( T \right),es\left( T \right)} \right)
\end{array}
\end{equation}
where $es\left( T \right)={{\left( v\left( T \right),h\left( T \right) \right)}^{T}}$; ${{W}_{ESE}}$, ${{W}_{FG}}$ represents the weights of explicit state encoder and fusion gate mechanism, respectively. $es\left( T \right)$, $s\left( T \right)$ represent the explicit state encoding and Koopman state approximation of the vehicle at time step $T$.
\vspace{1em}
\subsubsection{Koopman evolution block and decoder}
In the Koopman evolution block, the multi-step evolution in the observable high-dimensional linear system is achieved based on the Koopman operator approximation $K$ learned through two linear layers without bias. Specifically, $K$ comprises the system matrix $A$ and the control matrix $B$. $A$ describes the process of system state transition without control inputs, while the velocity of preceding vehicle in the prediction horizon is treated as the system control input. This input is applied to the system through the control matrix $B$ to complete the system evolution. The formal expression of this process is as follows:
\begin{equation}
\begin{array}{c}
{s^P}\left( {T + F} \right) = A{s^P}\left( {T + F - 1} \right) + B{v_{ - 1}}\left( {T + F - 1} \right)\\
 = A_{}^Fs\left( T \right) + \sum\limits_{f = 1}^F {A_{}^{f - 1}B{v_{ - 1}}\left( {T + F - f} \right)} \\
 \buildrel \Delta \over = \tilde A_{}^F\left( {s\left( T \right),{v_{ - 1}}\left( {T:T + F - 1} \right)} \right)
\end{array}
\end{equation}
where ${{s}^{P}}\left( \cdot  \right)$, ${{v}_{-1}}\left( \cdot  \right)$ represent the predicted state in high-dimensional space and the velocity of the preceding vehicle, respectively.

Subsequently, the decoder, characterized as a bias-free linear layer, serves to approximate the Koopman modes and reconstruct the predicted state from the observable high-dimensional space to the original space. To minimize online computation delay for CAVs, the decoder is specifically designed as a linear layer without bias (In Section 5, the predictive performance difference between the current decoder and using an MLP as the decoder will be demonstrated). The reconstructed state variables in the original space align with the state variables input to the explicit state encoder in AdapKoopnet. The reconstruction process is expressed as follows:
\begin{equation}
e{s^P}\left( {T + f} \right) = {W_{DEC}} \cdot {s^P}\left( {T + f} \right)
\end{equation}
where $e{{s}^{P}}\left( T+f \right)={{\left( {{v}^{P}}\left( T \right),{{h}^{P}}\left( T \right) \right)}^{T}}$ represents the predicted state of the original space; ${{W}_{DEC}}$ represents the learnable weights in the decoder.

\textbf{Remark 2}: In a typical trajectory prediction task, incorporating future velocities of the preceding vehicle as inputs may be impractical. However, the primary objective of AdapKoopnet is to predict the response driving behavior of HDVs to the preceding vehicle. This facilitates subsequent inferences about the required velocity of the CAVs. This makes this setup appear much more reasonable.

\subsection{Loss function}
Loss of AdapKoopnet is composed of reconstruction error, prediction error, and linear evolution error \citep{lusch2018deep}. Specifically, the reconstruction error represents the difference between the reconstructed state obtained by embedding the current explicit state of the vehicle into a high-dimensional space and reconstructing it through the decoder, and the original state. To achieve accurate reconstruction of the original state, the reconstruction error for the entire prediction horizon is included as part of the loss function, expressed as follows:
\begin{equation}
{L_C} = \frac{1}{{F + 1}}\sum\limits_{f = 0}^F {\left\| {{\varphi ^d}\left( {{\varphi ^e}\left( {es\left( {T + f} \right),x\left( {T - P + f:T + f} \right)} \right)} \right) - es\left( {T + f} \right)} \right\|} 
\end{equation}
where ${{\varphi ^d}}\left( \cdot  \right)$, ${{\varphi ^e}}\left( \cdot  \right)$ represent the transformations performed by the encoding block (highlighted by the yellow dashed box in Figure 2) and the decoder in AdapKoopnet; $\left\| \cdot  \right\|$ represents mean squared error.
The prediction error represents the difference between the predicted state and the ground truth, and it is defined as:

\begin{equation}
{L_P} = \frac{1}{F}\sum\limits_{f = 1}^F {\left\| {e{s^P}\left( {T + f} \right) - es\left( {T + f} \right)} \right\|} 
\end{equation}
The linear evolution error represents the difference between the Koopman state approximation at time $T$ after $F$ steps of linear evolution and the Koopman state approximation at time ${T+F}$. It is defined as:

\begin{equation}
{L_E} = \left\| {{s^P}\left( {T + F} \right) - s\left( {T + F} \right)} \right\|
\end{equation}
The loss function of AdapKoopnet is expressed as:
\begin{equation}
L = {\alpha _C}{L_C} + {\alpha _P}{L_P} + {\alpha _E}{L_E}
\end{equation}
where ${{\alpha }_{C}}$, ${{\alpha }_{P}}$, ${{\alpha }_{E}}$ are the weights corresponding to the three parts of the loss, and they are hyperparameters. To avoid manually selecting hyperparameters,  Dynamic Weight Aver age (DWA) are introduced \citep{liu2019end}. These weights are adjusted based on the losses from the previous epoch in the dynamic loss function.

\section{AdapKoopPC: Predictive control framework of CAVs for real-time optimizing mixed traffic flow based on Adakoopnet}
\subsection{Problem description and system state representation}

\begin{figure}[pos=htbp]
    \vspace{1em}  
    \centering
    \includegraphics[width=1\textwidth]{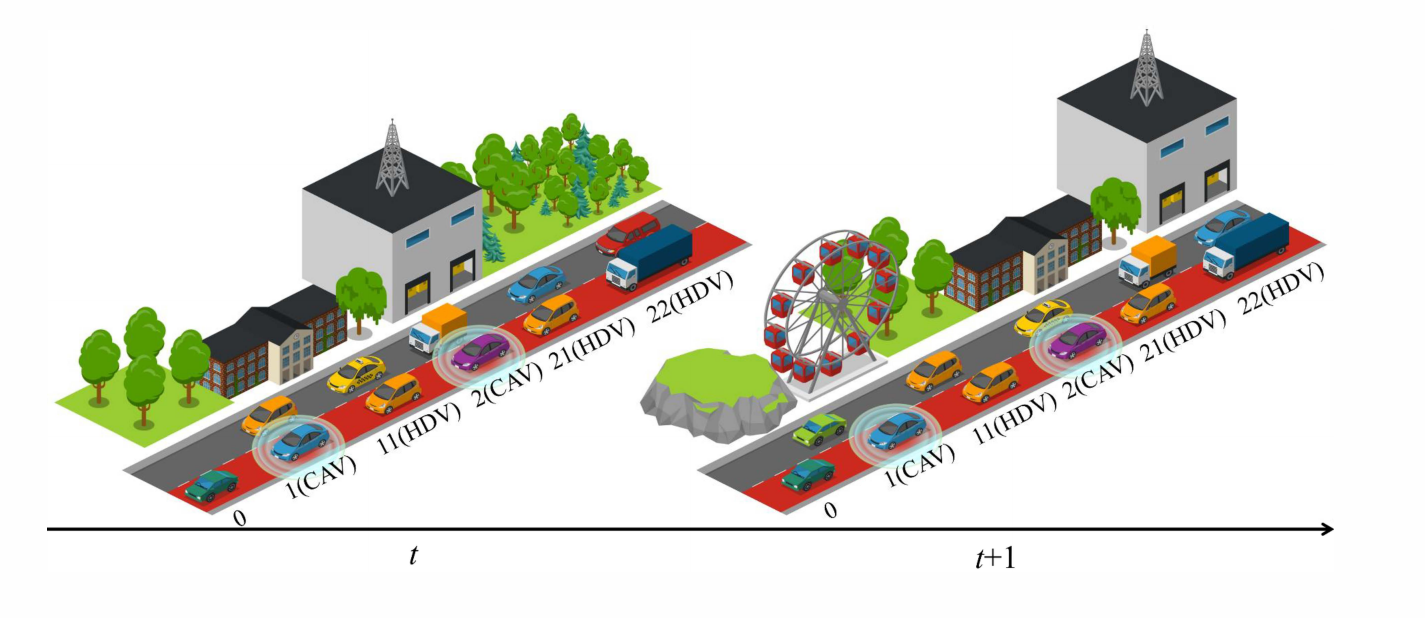}
    \vspace{-1em}
    \caption{\centering{The local mixed traffic system}}
	\label{Figure 14}
    \vspace{-1em}  
\end{figure}

Consider a local traffic system within a high-density mixed traffic flow, as illustrated in Fig. \ref{Figure 14} consisting of $N$ HDVs and $G$ CAVs driving in the same lane. The leading vehicle is a HDV numbered as $0$. For convenience, CAVs are sequentially numbered as ${1, 2, ..., G}$. The HDVs located between CAV $g$ and $g+1$ are sequentially numbered as $g1, g2, ..., g{n}_{g}$. The equivalence relationships exist: $\sum\limits_{g=1}^{G}{g{{n}_{g}}}=N$. Assuming that all HDV states in this system can be collected through the onboard sensors of CAVs or a V2X (Vehicle-to-Everything) system. The control problem of this system is defined as achieving the following objectives through the rolling optimization of control inputs for CAVs:
(1) Minimizing the velocity difference between any two adjacent vehicles in the system, while maintaining reasonable headway to alleviate traffic oscillations transmitted by the leading vehicle.
(2) Ensuring that the velocities of vehicles in the system are close to that of the leading vehicle, aiming to mitigate oscillations while maintaining traffic efficiency.

Based on the above two objectives and for the convenience of constraint formulation, the system state is defined as follows:
\begin{equation}
ES\left( t \right)=\left( {{p}_{1}}\left( t \right),{{v}_{1}}\left( t \right),{{a}_{1}}\left( t \right),{{h}_{11}}\left( t \right),{{v}_{11}}\left( t \right),\Delta {{v}_{11}}\left( t \right),...,{{h}_{G{{n}_{G}}}}\left( t \right),{{v}_{G{{n}_{G}}}}\left( t \right),\Delta {{v}_{G{{n}_{G}}}}\left( t \right) \right)
\end{equation}
where $ES\left( t \right)$ encompasses the headways, velocities of all vehicles in the system, as well as the velocity differences of HDVs and accelerations of CAVs; ${{p}_{1}}\left( t \right)$ is the position of the first CAV in the system. The control input of the system is defined as the jerk of CAVs, i.e., the rate of change of acceleration, denoted as:
\begin{equation}
U\left( t \right)=\left( {{u}_{1}}\left( t \right),{{u}_{2}}\left( t \right),...,{{u}_{G}}\left( t \right) \right)
\end{equation}

\textbf{Remark 3}: In large-scale high-density mixed traffic flow scenarios, the lane-changing frequency is low, areas without lane changes occur can be divided into multiple systems as described above. The system lacks a strict organizational structure, requiring cooperation among a few neighboring CAVs to establish. Unlike most existing research, the achievement of control objectives of the system does not depend on the so-called equilibrium velocity and equilibrium headway. Minimizing the velocity difference between adjacent vehicles is more readily accepted by HDV drivers, aligns with their own driving objectives in most cases. For scenarios with low penetration rates of V2X systems, simulation experiments relying solely on the onboard sensors of CAVs are conducted and tested in Section 6.

\subsection{State predictive model of mixed traffic system}
\begin{figure}[pos=htbp]
    \vspace{1em}  
    \centering
    \includegraphics[width=0.95\textwidth]{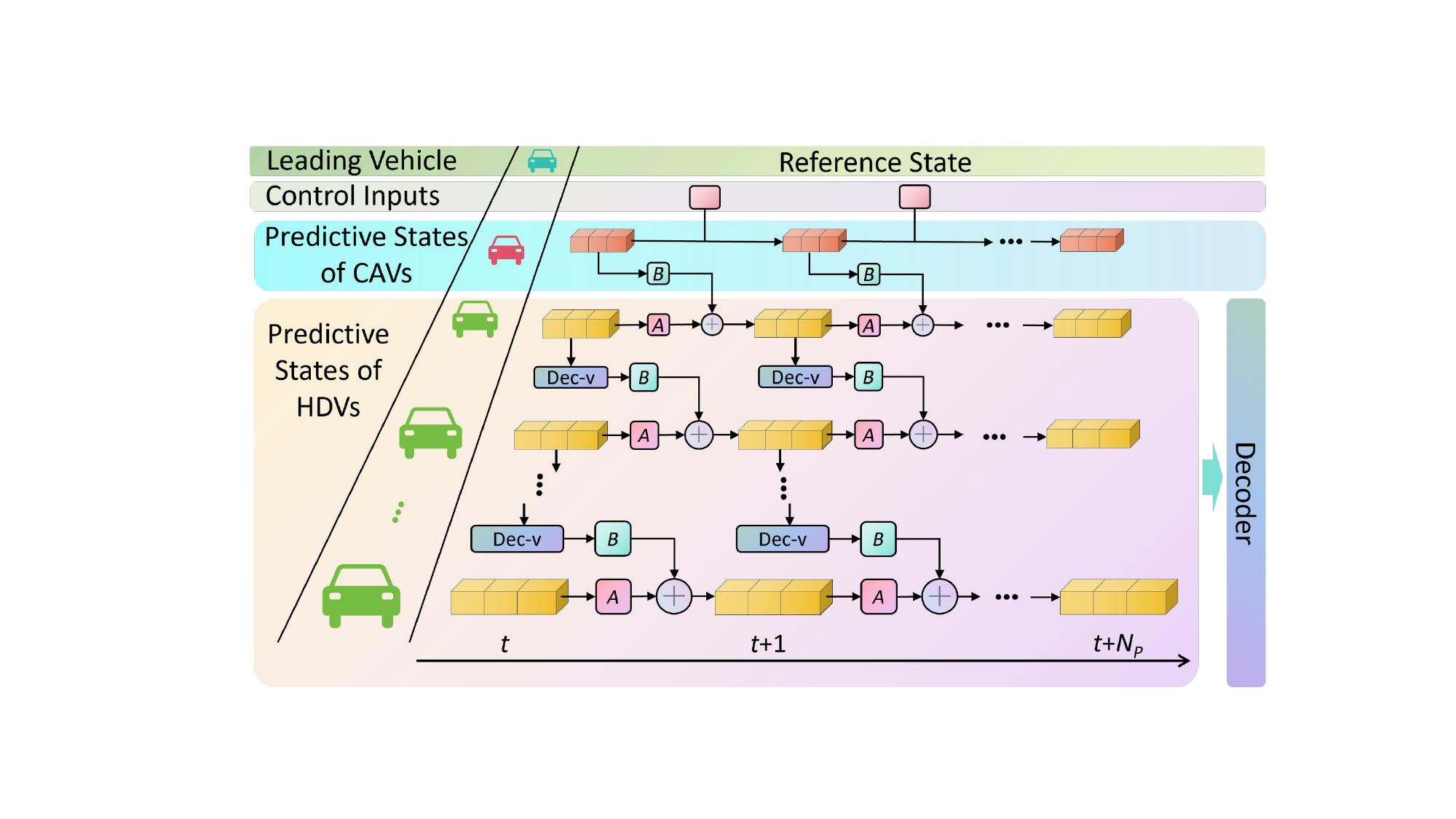}
    \vspace{-1em}
    \caption{\centering{State Predictive Process of Mixed Traffic System }}
	\label{Figure 4}
    \vspace{-1em}  
\end{figure}

To enable the rapid convergence of the system to the desired state while satisfying the constraints on system states and control inputs, a predictive control framework is adopted. The framework assumes a set of feasible control input sequences within a specified region known as the control horizon ${{N}_{C}}$. Based on the predictive model, a series of predictive states within the predictive horizon ${{N}_{P}}$ is obtained. By minimizing the error between the reference state and the predictive state, the optimal control input that satisfies constraints is determined. The current step control input is then applied to the CAVs, and this iterative process ensures that the control input at each step for the CAVs is at least a suboptimal solution. It is evident that in this framework, the accuracy of the system state predictive model is crucial. Firstly, the CAVs in the system are controllable, and their predictive state is obtained through the following vehicle kinematics equations:

\begin{equation}
\begin{array}{*{20}{l}}
{{p_q}\left( {t{\rm{ + }}1} \right){\rm{ = }}{p_q}\left( t \right){\rm{ + }}{v_q}\left( t \right) \cdot \Delta t}\\
{{v_q}\left( {t{\rm{ + }}1} \right){\rm{ = }}{v_q}\left( t \right){\rm{ + }}{a_q}\left( t \right) \cdot \Delta t}\\
{{a_q}\left( {t{\rm{ + }}1} \right){\rm{ = }}{a_q}\left( t \right){\rm{ + }}{u_q}\left( t \right) \cdot \Delta t}
\end{array} = \left[ {\begin{array}{*{20}{c}}
1&{\Delta t}&0\\
0&1&{\Delta t}\\
0&0&1
\end{array}} \right] \cdot \left[ {\begin{array}{*{20}{c}}
{{p_q}\left( t \right)}\\
{{v_q}\left( t \right)}\\
{{a_q}\left( t \right)}
\end{array}} \right] + \left[ {\begin{array}{*{20}{c}}
0\\
0\\
{\Delta t}
\end{array}} \right] \cdot {u_q}\left( t \right)
\end{equation}
where$\left[ \begin{matrix}
   1 & \Delta t & 0  \\
   0 & 1 & \Delta t  \\
   0 & 0 & 1  \\
\end{matrix} \right] \triangleq A_{CAV},\ \left[ \begin{matrix}
   0  \\
   0  \\
   \Delta t  \\
\end{matrix} \right] \triangleq B_{CAV}$  represent the system matrix and state matrix of CAVs, respectively. In this system, there is always at least one CAV in front of any HDV. Therefore, AdapKoopnet can be cascaded to predict the states of all vehicles in the system under specific control inputs of CAVs. This forms the basis for resolving traffic flow stop-and-go waves in the framework. The predictive process for all HDVs is illustrated in Fig. \ref{Figure 4}. Prior to the predictive process, it is necessary to obtain the high-dimensional representation of the current state of the vehicles using the encoding block of AdapKoopnet.

However, the predictive process seems to be cumbersome, particularly when system involves numerous vehicles. Therefore, the state predictive model needs to be integrated before initiating prediction to perform parallel prediction of all vehicle states within the system. For any HDV in system, the following relationship exists:
\begin{equation}
\left\{
\begin{aligned}
  & {{s}_{g{{n}_{g}}}}(t+1) = A \cdot {{s}_{g{{n}_{g}}}}(t) + B \cdot {{v}_{g{{n}_{g}}-1}}(t) \\ 
  & e{{s}_{g{{n}_{g}}}}(t) = C \cdot {{s}_{g{{n}_{g}}}}(t) \\ 
\end{aligned}
\right.
\end{equation}
where $C$ represents the observation matrix, which is the linear decoder in AdaptKoopnet. For the HDVs, where the preceding vehicle is also an HDV, the following equations are further obtained:
\begin{equation}
{s_{g{n_g}}}\left( {t + 1} \right) = A \cdot {s_{g{n_g}}}\left( t \right) + B \cdot {C_v} \cdot {s_{g{n_g}}}\left( t \right)
\end{equation}
where ${{C}_{v}}$ represents the row in the observation matrix used for observing velocity. Combining Eqs. (22)-(26), the predictive model for the system is obtained:
\begin{equation}
\left\{ \begin{array}{l}
S\left( {t + 1} \right) = {A_S} \cdot S\left( t \right) + {B_S} \cdot U\left( t \right)\\
ES\left( t \right) = {C_S} \cdot S\left( t \right)
\end{array} \right.
\end{equation}
where
\begin{equation}
\small
\begin{array}{l}
S\left( t \right) = {\left[ {{p_1}\left( t \right),{v_1}\left( t \right).{a_1}\left( t \right),{s_{11}}\left( t \right), \cdots ,{h_g}\left( t \right),{v_g}\left( t \right).\Delta {v_g}\left( t \right), \cdots ,{s_{g{n_g}}}\left( t \right), \cdots ,{s_{G{n_G}}}\left( t \right)} \right]^T}\\
{A_S} = \left[ {\begin{array}{*{20}{c}}
{{A_{CAV}}}&{}&{}&{}&{}&{}\\
B&A&{}&{}&{}&{}\\
{}&{B{C_v}}&A&{}&{}&{}\\
{}&{}& \ddots & \ddots &{}&{}\\
{}&{}&{}&{}&{{A_{CAV}}}&{}\\
{}&{}&{}&{}& \ddots & \ddots 
\end{array}} \right],{B_S} = \left[ {\begin{array}{*{20}{c}}
{{B_{CAV}}}&{}&{}\\
{}& \cdots &{}\\
{}&{}&{}\\
{}&{}&{{B_{CAV}}}\\
{}&{}&{}
\end{array}} \right],{C_S} = \left[ {\begin{array}{*{20}{c}}
{{I_3}}&{}&{}&{}&{}\\
{}&C&{}&{}&{}\\
{}&{}& \ddots &{}&{}\\
{}&{}&{}&{{I_3}}&{}\\
{}&{}&{}&{}& \ddots 
\end{array}} \right]
\end{array}
\end{equation}

\subsection{AdapKoopPC for mixed traffic system}
\subsubsection{Constrained optimization problem of mixed traffic system}
Assuming the predictive horizon is the same as the control horizon and denoting it as ${{N}_{P}}$, at time $t$, the optimization objective for the mixed traffic system defined in Section 4.1 can be formulated as minimizing the following cost function over the horizon to find an optimal set of control inputs ${{\mathrm{U}}^{*}}\left( t \right)={{\left[ {{U}^{*}}\left( \left. 0 \right|t \right),{{U}^{*}}\left( \left. 1 \right|t \right),\ldots ,{{U}^{*}}\left( \left. {{N}_{P}}-1 \right|t \right) \right]}^{T}}$:
\begin{equation}
J = \sum\limits_{i = 1}^{{N_P}} {\left\| {ES\left( {\left. i \right|t} \right) - E{S_{ref}}\left( {t + i} \right)} \right\|_Q^2 + \left\| {U\left( {\left. {i - 1} \right|t} \right)} \right\|_R^2} 
\end{equation}
where $ES\left( \left. i \right|t \right)$ is the predictive state of the system after time steps $i$ at time $t$; $E{{S}_{ref}}\left( t \right)$ represents the reference state, it provides the reference values for the velocity of CAVs in the system, namely the average velocities of their respective preceding vehicles over the time interval $t-{{N}_{P}}$ to $t$; $Q$ and $R = {r_u} \cdot {I_G}$ are diagonal matrices, representing the penalty weights for state and control input, respectively, where $Q$ contains penalties for velocities of CAVs and velocity differences of HDVs, defined as follows:

\begin{equation}
Q = \left[ {\begin{array}{*{20}{c}}
{{q_{CAV}}}&{}&{}&{}\\
{}&{{q_{HDV}}}&{}&{}\\
{}&{}& \ddots &{}\\
{}&{}&{}&{{q_{HDV}}}
\end{array}} \right],\;{q_{CAV}} = \left[ {\begin{array}{*{20}{c}}
0&{}&{}\\
{}&{q_{CAV}^v}&{}\\
{}&{}&0
\end{array}} \right],\;{q_{HDV}} = \left[ {\begin{array}{*{20}{c}}
0&{}&{}\\
{}&0&{}\\
{}&{}&{q_{HDV}^{\Delta v}}
\end{array}} \right]
\end{equation}
Additionally, the system states and control inputs must satisfy the following constraints:
\begin{equation}
\begin{array}{l}
{h_{\min }} < h\left( t \right) < {h_{\max }}\\
{v_{\min }} < v\left( t \right) < {v_{\max }}\\
{a_{\min }} < {a_{CAV}}\left( t \right) < {a_{\max }}\\
{u_{\min }} < u\left( t \right) < {u_{\max }}
\end{array}
\end{equation}
These constraints respectively imply that the headways of all vehicles in the system should remain within a reasonable range, velocities should stay within velocity limits, and the acceleration and control inputs of CAVs should adhere to vehicle dynamics constraints.

\subsubsection{Standard quadratic programming of AdapKoopPC}
To facilitate solving, the constrained optimization problem of mixed traffic system can be reformulated as a standard quadratic programming problem. Building on the linear model described in Eq.(25), the predictive states among the future ${{N}_{P}}$ steps can be expressed as:
\begin{equation}
\bm{\mathit{ES}}(t) = \bm{\mathit{AS}}(t) + \bm{\mathit{BU}}(t)
\end{equation}
where $\bm{\mathit{ES}}(t)={{\left[ ES{{\left( \left. 1 \right|t \right)}^{T}},\ldots ,ES{{\left( \left. {{N}_{P}} \right|t \right)}^{T}} \right]}^{T}}$; $\bm{\mathit{BU}}(t)={{\left[ U{{\left( \left. 0 \right|t \right)}^{T}},\ldots ,U{{\left( \left. {{N}_{P}}-1 \right|t \right)}^{T}} \right]}^{T}}$; $\bm{\mathit{A}}={{\left[ \begin{matrix}
   C{{A}_{S}}  \\
   \vdots   \\
   CA_{S}^{{{N}_{P}}}  \\
\end{matrix} \right]}^{T}},\ \bm{\mathit{B}}={{\left[ \begin{matrix}
   {{C}_{S}}{{B}_{S}} & {} & {} & {}  \\
   {{C}_{S}}{{A}_{S}}{{B}_{S}} & {{C}_{S}}{{B}_{S}} & {} & {}  \\
   \vdots  & \vdots  & \ddots  & {}  \\
   {{C}_{S}}A_{S}^{{{N}_{P}}-1}B & {{C}_{S}}A_{S}^{{{N}_{P}}-2}B & \cdots  & {{C}_{S}}{{B}_{S}}  \\
\end{matrix} \right]}^{T}}$. 

Let $\bm{\mathit{E}}{\bm{\mathit{S}}_{\mathrm{ref}}}\left( t \right)={{\left[ E{{S}_{ref}}{{\left( \left. 1 \right|t \right)}^{T}},\ldots ,E{{S}_{ref}}{{\left( \left. {{N}_{P}} \right|t \right)}^{T}} \right]}^{T}}$, $\bm{\mathit{Q}}={{I}_{{{N}_{P}}}}\otimes Q$, $\bm{\mathit{R}}={{I}_{{{N}_{P}}}}\otimes R$, where $\otimes $ is Kronecker product. And by discarding the terms without $\bm{\mathit{U}}(t)$, the cost function Eq.(29) has been rewritten as follows:
\begin{equation}
J=\frac{1}{2}\bm{\mathit{U}}{{\left( t \right)}^{T}}H\bm{\mathit{U}}\left( t \right)+{{F}^{T}}\bm{\mathit{U}}\left( t \right)
\end{equation}
where $H=2\left( {{\bm{\mathit{B}}}^{T}}\bm{\mathit{QB}}+\bm{\mathit{R}} \right)$; ${{F}^{T}}=2{{\left( \bm{\mathit{A}}S\left( t \right)-\bm{\mathit{E}}{{\bm{\mathit{S}}}_{\mathrm{ref}}}\left( t \right) \right)}^{T}}\bm{\mathit{QB}}$. Optimal control input sequence can be obtained by solving the constrained optimization problem (33) subject to Eqs. (31) and (32).

\section{Experiment Part I: AdapKoopnet for car following trajectory prediction of HDVs }

\subsection{Dataset description and evaluation metrics}

The performance of AdapKoopnet model is evaluated on the HighD dataset, which is a large-scale naturalistic vehicle trajectory dataset collected from German highways. The dataset comprises 11.5 hours of measurements and covers 110,000 vehicles, with a total measured vehicle distance of 45,000 km, capturing trajectories of both passenger cars and trucks. The car-following trajectory dataset is extracted in accordance with the criteria presented in \citep{mo2021physics}, includes eleven fields such as id, precedingId, xVelocity, xAcceleration, and precedingXVelocity, totaling 18396432 records. And the root mean square error (RMSE) is selected as the evaluation metrics:
\begin{equation}
RMSE = \sqrt {\frac{1}{{\Gamma }}\sum\limits_{i = 1}^{\Gamma } {{{({y_i} - {{\bar y}_i})}^2}} } 
\end{equation}

\subsection{Baseline models and experiment settings}
The performance of designed AdapKoopnet for HDVs car following trajectory Prediction is compared and evaluated with the following baseline models.

\noindent \textbf{MLP} utilizes a multi-layered nonlinear fully connected neural network to predict the car-following trajectory of vehicle.

\noindent \textbf{LSTM} utilizes LSTM-based encoder-decoder architecture to predict vehicle trajectories \citep{park2018sequence}.

\noindent \textbf{Koopnet} is an ablation version of AdapKoopnet where the extraction of driving characteristics semantic information from the trajectories context of vehicle is eliminated. This emphasizes the importance of the driving characteristic extraction block.

\noindent \textbf{N-AdapKoopnet} is a variant of AdapKoopnet where the decoder is replaced with fully connected layers featuring non-linear activation functions. This demonstrates the impact of linear decoder on the prediction results.

\noindent \textbf{N-Koopnet} is similar to N-AdapKoopnet, features an encoder composed of fully connected layers with non-linear activation functions.

\noindent \textbf{S-AdapKoopnet} employs a modified architecture in which the width of each layer is halved compared to AdapKoopnet while retaining the same structure. This modification aims to reduce the computational load, especially when implementing optimization for large-scale mixed traffic systems.

PyTorch 2.1.0 framework is utilized to construct the predictive model, and end-to-end training was performed under a platform with Intel Core i9-13900K processor and NVIDIA RTX 4090 GPU. The processed car-following trajectories were split into training, validation, and test sets in a ratio of 7:1:2. The model relevant hyperparameters are detailed in \ref{Tabel:1}.

\begin{table}[pos=h]
  \caption{Model hyperparameter}\label{Tabel:1}
  \centering
 \setlength{\tabcolsep}{10mm}{
\begin{tabular}{|c|c|c|}
\hline
Hyperparameter Type                  & Hyperparameter               & Values \\ 
\hline
                                    & Number of Layers & 3  \\          
Explicit State Encoder	 & MLP activation function & Tanh \\
  &  Dropout	                        & 0.2     \\ 
\hline
                                    & Trajectory Context horizon & 31	  \\
                                    & Driving Scenario type & 3  \\            
Semantic Extraction Block	 & Attention head & 4 \\
  &  Attention dimension	                 & 64     \\  
&  Input feature dimensione                        & 5     \\ 
\hline
                                    & Model dimension & 128   \\
                                    & Prediction horizon & 15  \\             
                                    & Batch size  & 256  \\            
Other hyperparameters	 & Max train epochs & 25 \\
&  learning rate(LR)                      & $10^{-5}$     \\  
&  LR schedular                   & Exponent      \\ 
&  LR decay rate                   & 0.6     \\ 
\hline
\end{tabular}
}
\end{table}

\subsection{Trajectory prediction experiment results}
\subsubsection{HDVs trajectory prediction performance comparison}
Table \ref{Tabel:2} statistics the performance index results of AdapKoopNet and other baseline models for trajectory prediction under different prediction horizons, including the RMSE of velocity and headway. In terms of velocity, when the prediction step is 5 (0.6 seconds), the RMSE of LSTM is the smallest at 0.084m. The reason may be due to the architectural design of the LSTM encoder-decoder and the excellent gate control unit structure, which has advantages in short time domain prediction. When the prediction step increases to 1.2 seconds, AdapKoopNet has a significant advantage, with an RMSE of 0.145m, which means it has better long-term prediction ability. AdapKoopnet considers the multi-step evolution loss of linear space during the training process, giving up some short-term prediction performance in exchange for the average prediction performance in the entire prediction time domain. It is also worth mentioning that although Koopman theory provides a global linear expression of a dynamic system in an infinite-dimensional space, AdapKoopnet ultimately implements predictions in a finite-dimensional (128-dimensional) space, which is bound to be affected by linear systems, resulting in some losses. Simply put, with the help of the powerful feature extraction capability of the attention mechanism and the adaptive driving feature extraction architecture based on driving scenarios, the AdapKoopnet series models have achieved performance comparable to LSTM in terms of short-term prediction performance, and better than it in terms of long-term prediction performance.

Although as the prediction step further increases, N-AdapKoopNet outperforms AdapKoopNet in performance, reflecting the loss of prediction performance caused by the adoption of linear decoders, the average performance at each prediction step size is still the most outstanding for AdapKoopNet. Subsequently, comparing the distance between the front of the vehicle, the most effective models are LSTM and N-AdapKoopNet. However, the decoder of this model is non-linear and can indeed achieve good trajectory prediction results. However, it is also not suitable for subsequent CAVs prediction control. Taking into account the accuracy of trajectory prediction and the feasibility of subsequent control for real-time traffic flow optimization, it is evident that AdapKoopNet has significant advantages. In addition, in order to balance prediction performance and inference cost, the large-scale hybrid transportation system simulation experiment in Section 6.3 is based on S-AdapKoopnet.

\begin{table}[pos=h]
\caption{Comparison of HDVs longitudinal trajectory prediction performance indicators}\label{Tabel:2}
  \centering
\setlength{\tabcolsep}{3.5mm}{
\begin{tabular}{@{} |c|cccc|cccc|@{} }
\hline
\multirow{2}{*}{Model} & \multicolumn{4}{c|}{Velocity RMSE(m/s)} & \multicolumn{4}{c|}{Headway RMSE (m)} \\
\cline{2-9} 
         & 0.6s    & 1.2s   & 1.8s    & Average    & 0.6s    & 1.2s   & 1.8s    & Average    \\
         \hline
MLP           & 0.184     & 0.341 & 0.472 & 0.3324    & 0.447 & 0.659      & 0.882  & 0.663 \\
LSTM    & \textbf{0.084}    & 0.173 & 0.367 & 0.208    & \textbf{0.444} & \textbf{0.628}     & 0.797  & 0.623 \\
Koopnet    & 0.176     & 0.318 & 0.444 &0.313   & 0.468 & 0.672   & 0.883  & 0.674 \\
N-AdapKoopnet     & 0.110     & 0.155& \textbf{0.243}& 0.169  & 0.455 &\textbf{0.628}      & \textbf{0.769}  & \textbf{0.618} \\
N-Koopnet      & 0.200    & 0.322 & 0.444 & 0.322   & 0.454& 0.647      & 0.851 & 0.650 \\
S-AdapKoopnet      & 0.105     & 0.148 & 0.247 & 0.167    & 0.477 & 0.651      & 0.796 & 0.641 \\
\hline \hline
AdapKoopnet     &  0.095  &  \textbf{0.145}    & 0.2471&  \textbf{0.162} &  0.466 & 0.643     & 0.792 & 0.634\\
\hline
\end{tabular}
}
\end{table}

\begin{figure}[pos=htbp]
    \vspace{1em}  
    \centering
    \includegraphics[width=1\textwidth]{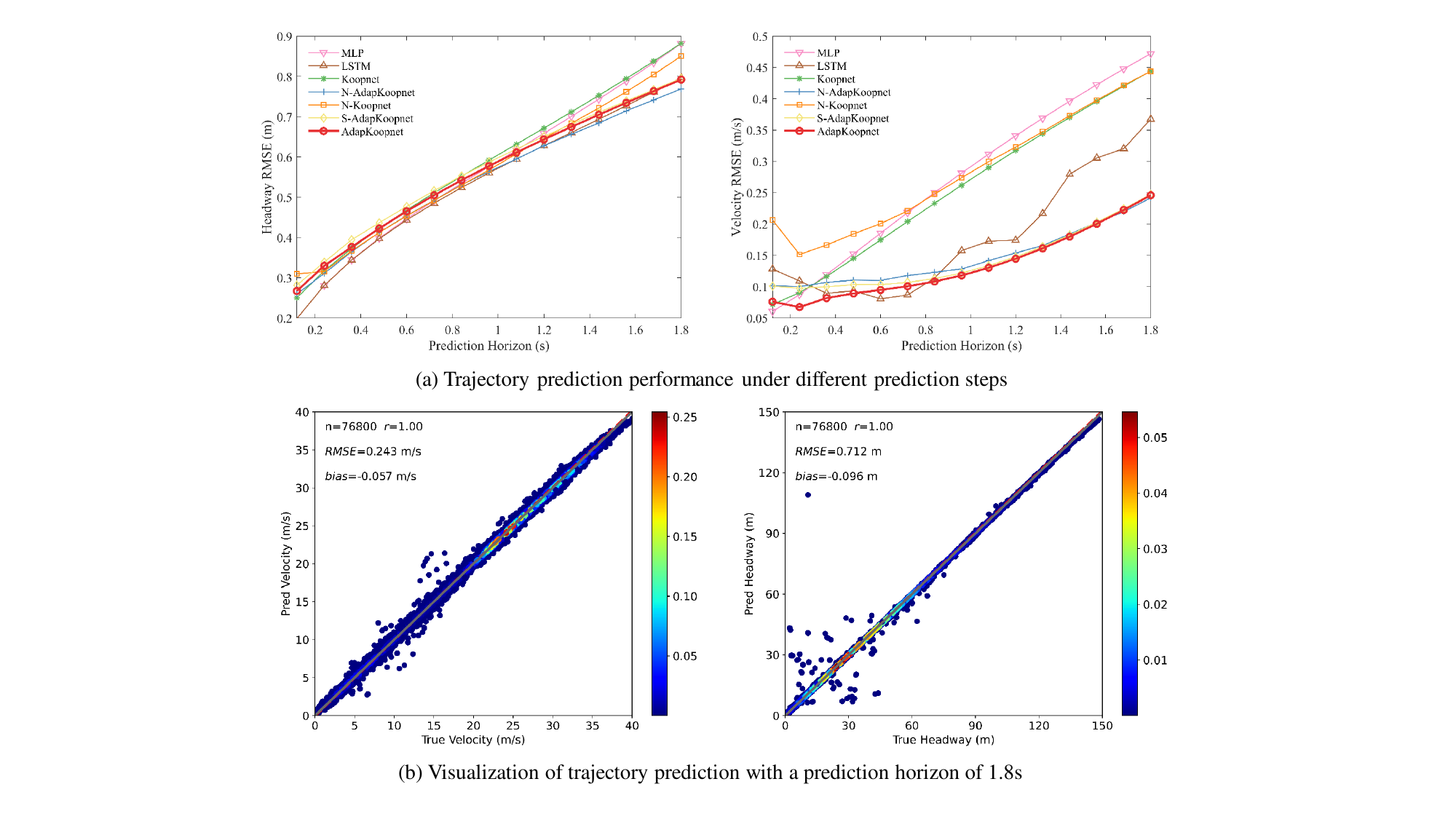}
    \vspace{-1em}
    \caption{\centering{HDVs driving state prediction performance of the AdapKoopnet }}
	\label{Figure 5}
    \vspace{-1em}  
\end{figure}

Fig. \ref{Figure 5} (a) shows the indicator line graph of AdapKoopNet and the baseline models under  complete 15 prediction steps, where the red line represents AdapKoopNet. The comparison results are obvious, mainly including the following two aspects. Firstly, regardless of the prediction range, the AdapKoopNet model outperforms KoopNet, indicating that it effectively and accurately captures the underlying data features and achieving personalized driving behavior prediction. Secondly, as the prediction range increases, although the prediction performance of AdapKoopNet has declined, it still maintains accuracy and feasibility. The comparative analysis with the baseline models is similar to Table \ref{Tabel:2} and will not be further elaborated.

Fig. \ref{Figure 5} (b) focuses specifically on the prediction step scene of 1.8s. 300 batches are randomly selected from the test set, and the predicted and true values of headway and velocity are visualized, with the color axis representing the probability density. It can be observed that the velocity of most trajectory appears between 20 and 30 meters, while the headway remains around 30 to 50 meters. In addition, the vast majority of  trajectory samples are attached near the diagonal, which means that the predicted values are very close to the true values, reflecting the strong ability of AdapKoopNet to capture trajectory evolution features and make accurate predictions. Of course, there are some samples with significant deviations, but it can be seen that these scattered points are all dark blue, indicating that the density of these samples is extremely low and almost does not affect the overall trajectory prediction performance. Of course, benefiting from the rolling optimization method adopted in the AdapKoopPC, the negative impact of these deviation samples on the subsequent mixed traffic flow optimization control can be ignored.

\subsubsection{Potential scenarios recognition and clustering results}
Furthermore, in order to verify the effectiveness of AdapKoopnet in directly extracting potential driving scenarios from real trajectory data and adaptively learning driving characteristics without any pre-define or pre-label, a series of visualizations of the model training results are performed.

\begin{figure}[pos=htbp]
    \vspace{1em}  
    \centering
    \includegraphics[width=1\textwidth]{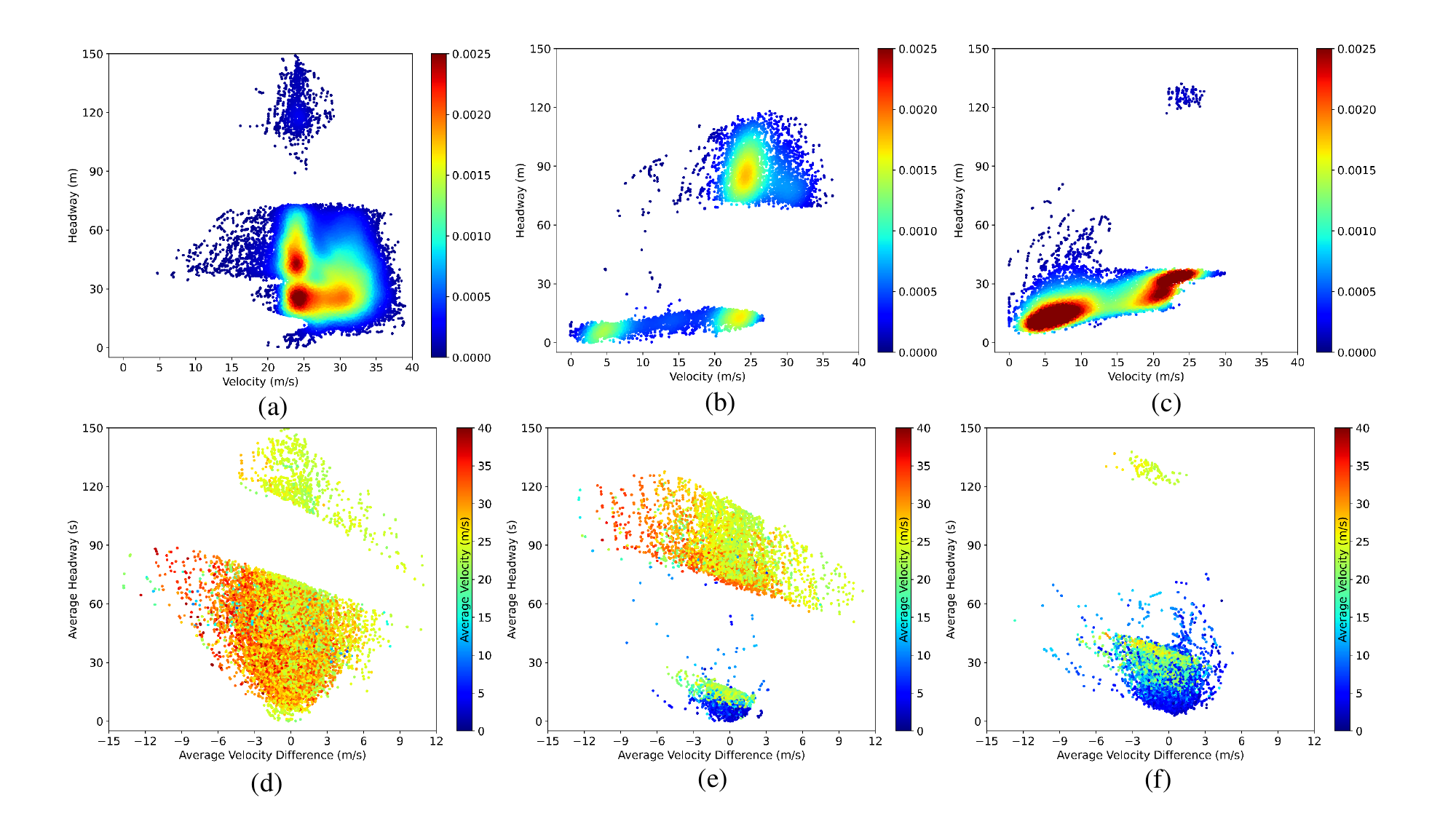}
    \vspace{-1em}
    \caption{\centering{Adaptive Multi-driving scenario recognition results}}
	\label{Figure 6}
    \vspace{-1em}  
\end{figure}

The first thing to explore is what all trajectory samples output after passing through the multi-head driving scenario recognition module. Fig. \ref{Figure 6} shows the distribution of trajectory samples for various driving scenarios learned and clustered. Fig. \ref{Figure 6} (a), (b) and (c)show the headway-velocity relationship of trajectory samples under three driving scenarios. The color axis represents the proportion of samples, with the proportion of trajectory samples increasing as the color approaches red. It is obvious that Fig. \ref{Figure 6} (a) tends towards medium to high velocity driving scenario, with the highest proportion of sample clusters distributed at velocity of around 25m/s, accompanied by a moderate headway of 30 meters to 60 meters. The distribution of trajectory samples in Fig. \ref{Figure 6} (b) is relatively uniform, including scenario with small headway at low velocity and large headway at high velocity. Fig. \ref{Figure 6} (c) corresponds to the third driving scenario, where the overall trajectory exhibits low velocity accompanied by small headway. Fig. \ref{Figure 6} (d), (e), and (f) depicts the relationship between the average velocity difference, average headway, and average velocity of trajectory samples in three driving scenarios, with the color axis representing velocity. Significantly, all trajectory samples are clearly clustered into three driving scenarios, each with similar repetitive patterns. Of course, the characteristics of all trajectories in each posture scenario are not entirely the same. Overall, AdapKoopNet has the ability to adaptively extract features from a large number of trajectory samples and cluster them, ultimately forming these three potential driving scenarios. 

\subsubsection{Scenario inherent temporal correlation and the driving characteristics dynamic temporal correlation}
\begin{figure}[pos=htbp]
    \vspace{1em}  
    \centering
    \includegraphics[width=1\textwidth]{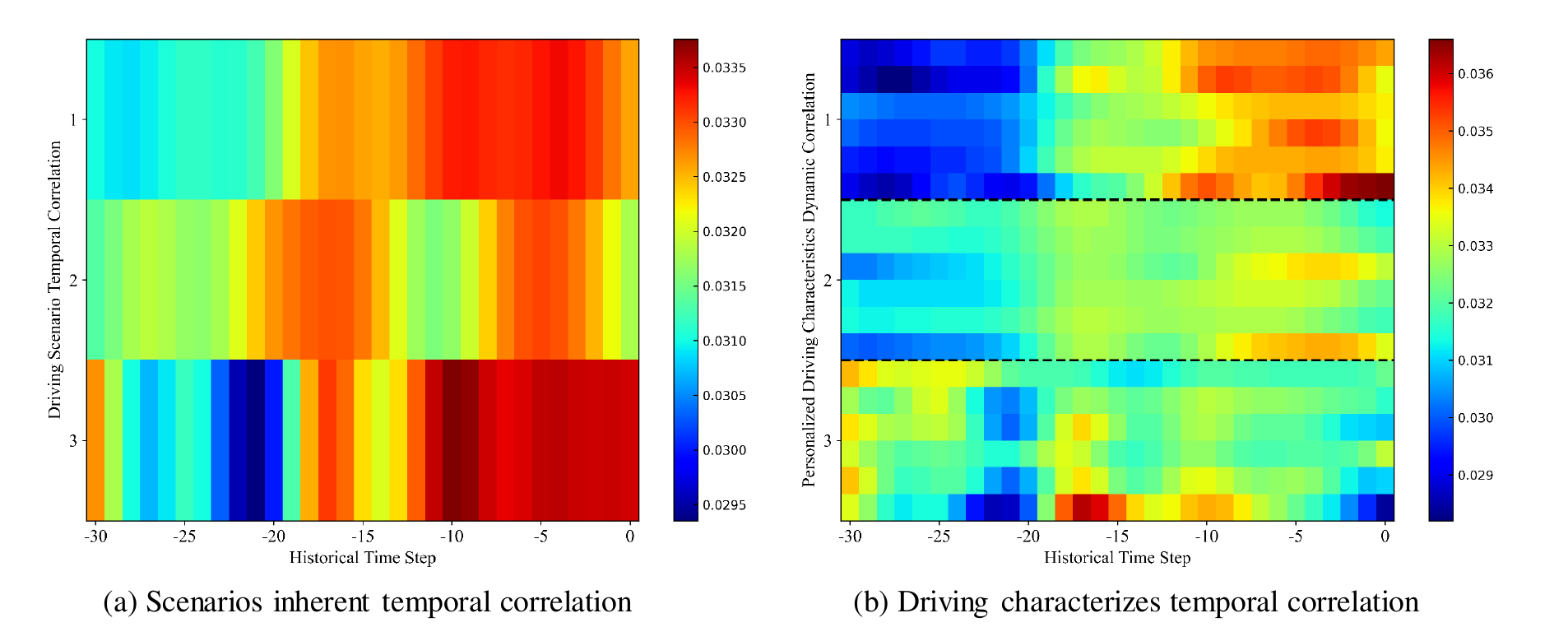}
    \vspace{-1em}
    \caption{\centering{The inherent temporal correlation of driving scenarios and the temporal correlation of driving characteristics}}
	\label{Figure 7}
    \vspace{-1em}  
\end{figure}

Fig. \ref{Figure 7} reveals this specific driving scenario inherent temporal correlation and the driving characteristics dynamic temporal correlation. Fig. \ref{Figure 7} (a) corresponds to the output of the driving characteristics semantic module shown in \ref{Figure 3} (b) in AdapKoopNet. It is worth noting that as mentioned earlier, AdapKoopNet adaptively learns the feature weights of each trajectory sample belonging to three different scenarios, and then clusters them into potential scenarios according to their propensity. Therefore, more precisely, what is extracted here is the inherent temporal correlation corresponding to each driving scenario. Specifically, driving scenario 1 and driving scenario 3 are quite similar, both reflecting an overall trend of lower temporal correlation as the historical time step increases. The difference lies in that the former is not as extreme as the latter, and Scenario 3 strongly relies on the nearest time step and has almost no correlation on distant historical trajectories. There is a significant difference between driving scenario 2 and the above two driving scenarios, characterized by a more stable temporal correlation on the historical trajectory of each time step, rather than being more affected as time approaches.

Fig. \ref{Figure 7} (b) indicates the dynamic correlation that directly reflects the actual driving characteristics of each vehicle. Here, we visualized six vehicles trajectories in each scenario. Significantly, whether it is driving scenarios 1, 2, or 3, the corresponding six trajectories exhibit dynamic correlations equivalent to the six variants in each driving scenario. They tend towards a scenario and follow the inherent temporal characteristics of the driving scenario, but the driving characteristics of each trajectory are different from each other, corresponding to the driving characteristic dynamic correlation module. These indeed demonstrate that AdapKoopNet has adaptively learned the inherent temporal correlations and driving characteristics dynamic correlations in different driving scenarios. The capture of these correlations contributes to subsequent accurate and personalized trajectory prediction, which is elaborated and proven in the following section.

\subsubsection{Validation of scenarios recognition effectiveness}
\begin{figure}[pos=htbp]
    \vspace{1em}  
    \centering
    \includegraphics[width=1\textwidth]{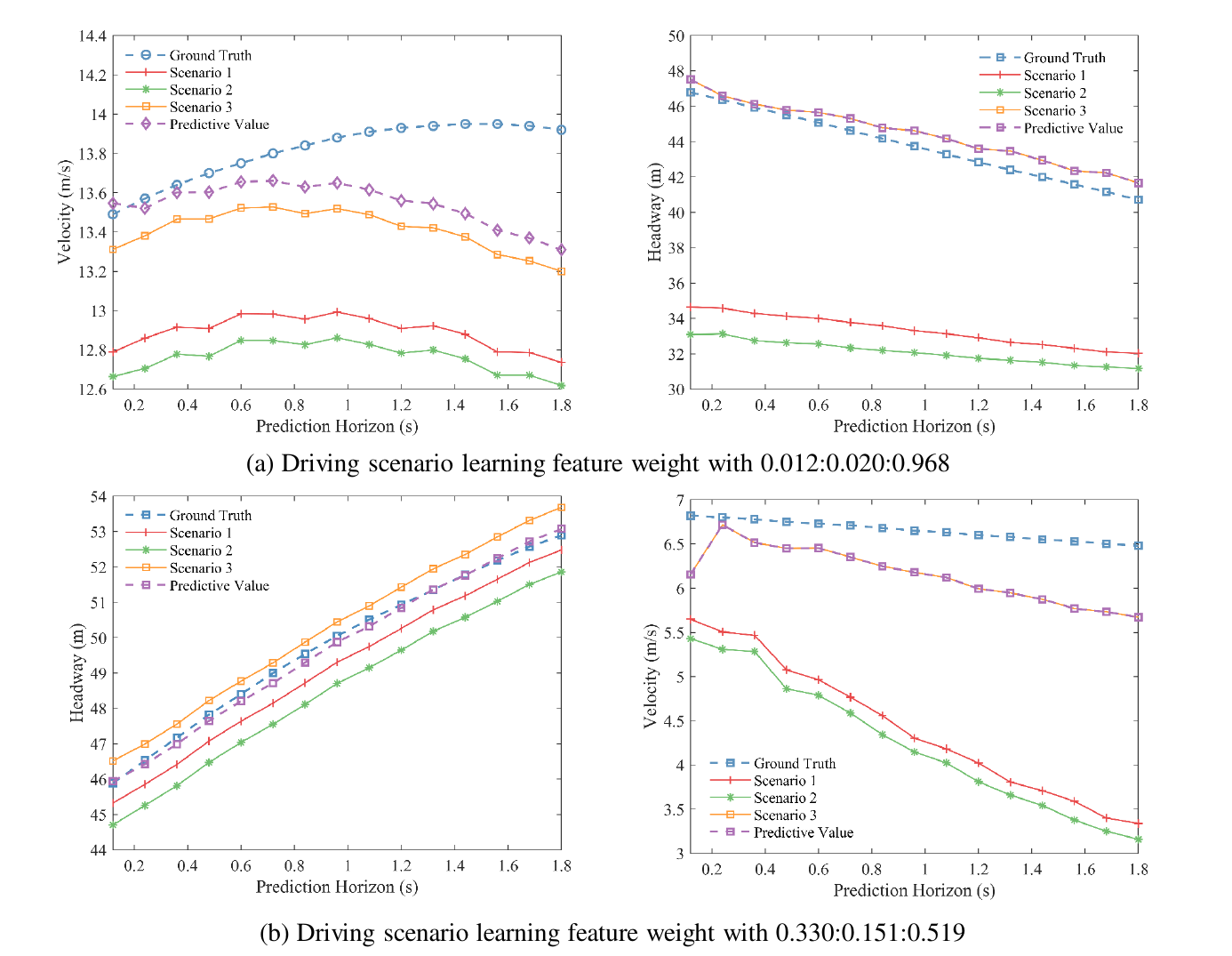}
    \vspace{-1em}
    \caption{\centering{Trajectory prediction results for the same trajectory under different driving scenario}}
	\label{Figure 8}
    \vspace{-1em}  
\end{figure}

Fig. \ref{Figure 8} is a comparative experiment demonstrating the effectiveness of the multi-head driving scenario recognition module and the driving characteristics semantic transformation module in learning potential driving scenarios and dynamic temporal correlation. Two trajectory test samples were selected, corresponding to adaptive learning of three driving scenario characteristic weight of 0.012:0.020:0.968 and 0.330:0.151:0.519, corresponding to Fig. \ref{Figure 8} (a) and Fig. \ref{Figure 8} (b), respectively. A very direct and convincing approach is to manually input different driving scenario labels for the same trajectory, and then compare its trajectory prediction results with those of unlabeled adaptive learning. Firstly, as shown in  Fig. \ref{Figure 8} (a), the feature weight of driving scenario 3 is 0.968, which is almost close to 1. The predicted trajectory with manually labeled scenario 3 is almost completely consistent with the trajectory prediction results with unlabeled adaptive learning, while the trajectory prediction results in manually labeled driving scenarios 2 and 3 are significantly different. The predicted velocity parameters of the trajectory show the same comparative results. By comparison, Fig. \ref{Figure 8} (b) shows that the feature weight of driving scenario 3 learned by AdapKoopNet is 0.519, and the trajectory features are more inclined towards scenario 3. It is obvious that in both headway and velocity aspects, the trajectory prediction results for scenario 3 with manual label driving represented by the yellow line, closely align with the observed values, followed by scenario 1 with manual label, and scenario 2 with manual label has the largest trajectory prediction deviation. The above comparison serves as strong evidence of the effectiveness and superiority of AdapKoopNet. In addition, there is actually another discovery that as the prediction step size increases, the prediction trends under various preset conditions are similar in trajectory prediction. This phenomenon arises from the fact that different prediction conditions correspond to different Koopman operators, with the only variation being the associated feature vectors. This is worth further improvement in future research.

\section{Experiment Part II: AdapKoopPC for mixed traffic system }

This section aims to comprehensively evaluate the effectiveness and superiority of the proposed AdapKoopPC in mitigating traffic oscillations in mixed traffic flows. Specifically, it covers small-scale experiments and large-scale experiments. The former aims to compare the spatio-temporal evolution trajectory of each vehicle in the mixed traffic system based on AdapKoopPC and based on other benchmark control methods, see Section 6.2; the latter is to verify the application of AdapKoopPC for the performance and generalization ability of large-scale mixed transportation systems, see Section 6.3.

\subsection{Simulation settings}
The small-scale simulated mixed traffic system consists of 2 CAVs, 2 HDVs(Truck) and 6 HDVs(Car). The large-scale simulated mixed traffic system consists of 50 vehicles, of which CAVs and HDVs(Car) account for 80\%, and HDVs(Truck) account for 20\%. A certain proportion of trucks is set up to build a heterogeneous simulation environment, with the purpose of proving the modeling and prediction capabilities of AdapKoopPC for heterogeneous HDVs. IDM is used as the simulation control model of HDVs, and its parameters are calibrated and obtained from the naturalistic driving dataset using genetic algorithm. The simulation is conducted on an open road, and the head vehicle of the mixed traffic system evolves according to the following equation to simulate traffic oscillations \citep{ruan2022impacts}: 

\begin{equation}
v_0(t) =
\left\{
\begin{array}{ll}
25\,\text{m/s} & t \in [0, 4.8]\,\text{s} \\
25 - 5\sin\left(0.1667(t - 4.8)\right)\,\text{m/s} & t \in [4.8, 180]\,\text{s}
\end{array}
\right.
\end{equation}

Except for the first vehicle, which must be a CAV, the arrangement of vehicles in the system is randomly generated based on the proportion of vehicle types. The control input for CAVs is solved using the SLSQP optimizer from the Scipy library, and the simulation is conducted on a laptop equipped with the Apple M2 chip. The simulation experiment parameters are shown in Table \ref{Tabel:3}.

\begin{table}[pos=h]
  \caption{Parameter settings for comparative experiments with different control methods}\label{Tabel:3}
  \centering
 \setlength{\tabcolsep}{4mm}{
\begin{tabular}{|c|cc|cc|cc|}
\hline
    & Parameter &Value  & Parameter &Value & Parameter &Value \\ 
\hline
            & $a_{IDM}^{car}\left( {{{\rm{m}} \mathord{\left/
 {\vphantom {{\rm{m}} {{{\rm{s}}^{\rm{2}}}}}} \right.
 \kern-\nulldelimiterspace} {{{\rm{s}}^{\rm{2}}}}}} \right)$ & 1.13 &   $a_{IDM}^{truck}\left( {{{\rm{m}} \mathord{\left/
 {\vphantom {{\rm{m}} {{{\rm{s}}^{\rm{2}}}}}} \right.
 \kern-\nulldelimiterspace} {{{\rm{s}}^{\rm{2}}}}}} \right)$ & 1.5 & $v_{IDM}^{car}\left( {{{\rm{m}} \mathord{\left/
 {\vphantom {{\rm{m}} {{{\rm{s}}^{\rm{2}}}}}} \right.
 \kern-\nulldelimiterspace} {{{\rm{s}}^{\rm{2}}}}}} \right)$ & 35.96\\          
IDM	 & $b_{IDM}^{car}\left( {{{\rm{m}} \mathord{\left/
 {\vphantom {{\rm{m}} {{{\rm{s}}^{\rm{2}}}}}} \right.
 \kern-\nulldelimiterspace} {{{\rm{s}}^{\rm{2}}}}}} \right)$ & 4  & $b_{IDM}^{truck}\left( {{{\rm{m}} \mathord{\left/
 {\vphantom {{\rm{m}} {{{\rm{s}}^{\rm{2}}}}}} \right.
 \kern-\nulldelimiterspace} {{{\rm{s}}^{\rm{2}}}}}} \right)$ & 4 & $v_{IDM}^{truck}\left( {{{\rm{m}} \mathord{\left/
 {\vphantom {{\rm{m}} {{{\rm{s}}^{\rm{2}}}}}} \right.
 \kern-\nulldelimiterspace} {{{\rm{s}}^{\rm{2}}}}}} \right)$ & 54.25  \\
  &  $s_0^{car}\left( {\rm{m}} \right)$ & 8.16  &$s_{IDM}^{truck}\left( {\rm{m}} \right)$ & 9.66& $l_{}^{car}\left( {\rm{m}} \right)$& 4.24   \\ 
  &  $T_0^{car}\left( {\rm{m}} \right)$ & 1.13 &$T_0^{truck}\left( {\rm{m}} \right)$ & 1.72& $l_{}^{truck}\left( {\rm{m}} \right)$& 11.82 \\ 
\hline
   &  ${h_{\min }}\left( {\rm{m}} \right)$ & 20 &${v_{\min }}\left( {\rm{m}} \right)$ & 0& ${a_{\max }}\left( {{{\rm{m}} \mathord{\left/
 {\vphantom {{\rm{m}} {{{\rm{s}}^{\rm{2}}}}}} \right.
 \kern-\nulldelimiterspace} {{{\rm{s}}^{\rm{2}}}}}} \right)$& 6 \\
AdapKoopPC &  ${h_{\max }}\left( {\rm{m}} \right)$ & 150 &${v_{\max }}\left( {\rm{m}} \right)$ & 150& ${a_{\min }}\left( {{{\rm{m}} \mathord{\left/ {\vphantom {{\rm{m}} {{{\rm{s}}^{\rm{2}}}}}} \right.
 \kern-\nulldelimiterspace} {{{\rm{s}}^{\rm{2}}}}}} \right)$& -6 \\
  &  ${u_{\min }}\left( {{\rm{m/}}{{\rm{s}}^3}} \right)$ & -6  &${u_{\max }}\left( {{\rm{m/}}{{\rm{s}}^3}} \right)$ & 6& ${N_P}$ & 10  \\ 
  &  $q_{HDV}^{\Delta v}$& 20 &$q_{CAV}^v$& 10 & $r_u$& 2 \\
\hline
Simulation & Duration(s) & 180  & Interval(s) & 0.12 &Truck proportion& 20\% \\
\hline
\end{tabular}
}
\end{table}

\subsection{Real-time control optimization experiment for small-scale mixed traffic system}
The Small-scale experiment aims to validate the effectiveness of AdapKoopnet in balancing  traffic flow disturbances and enhancing traffic flow stability, as well as the computational efficiency. The baseline control methods including: 

\noindent \textbf{LTI-MPC}: On the premise that the car-following model adopted by HDVs in the mixed traffic system is known, linear time-invariant MPC is used to optimize the mixed traffic system based on the linearized car-following model \citep{wang2023distributed}.

\noindent \textbf{Deep-LCC}: A data-driven non-parametric predictive control framework for mitigating traffic flow velocity oscillations. And the evolution dynamics of mixed traffic systems are  learned online using the pre-collected trajectories \citep{wang2023distributed}.

Additionally, to make a sound comparison of the proposed AdapKoopPC with the baselines, a fixed random seed is setting in the small-scale simulation.

\subsubsection{Mixed traffic system evolution results}
\begin{figure}[pos=htbp]
    \vspace{1em}  
    \centering
    \includegraphics[width=0.9\textwidth]{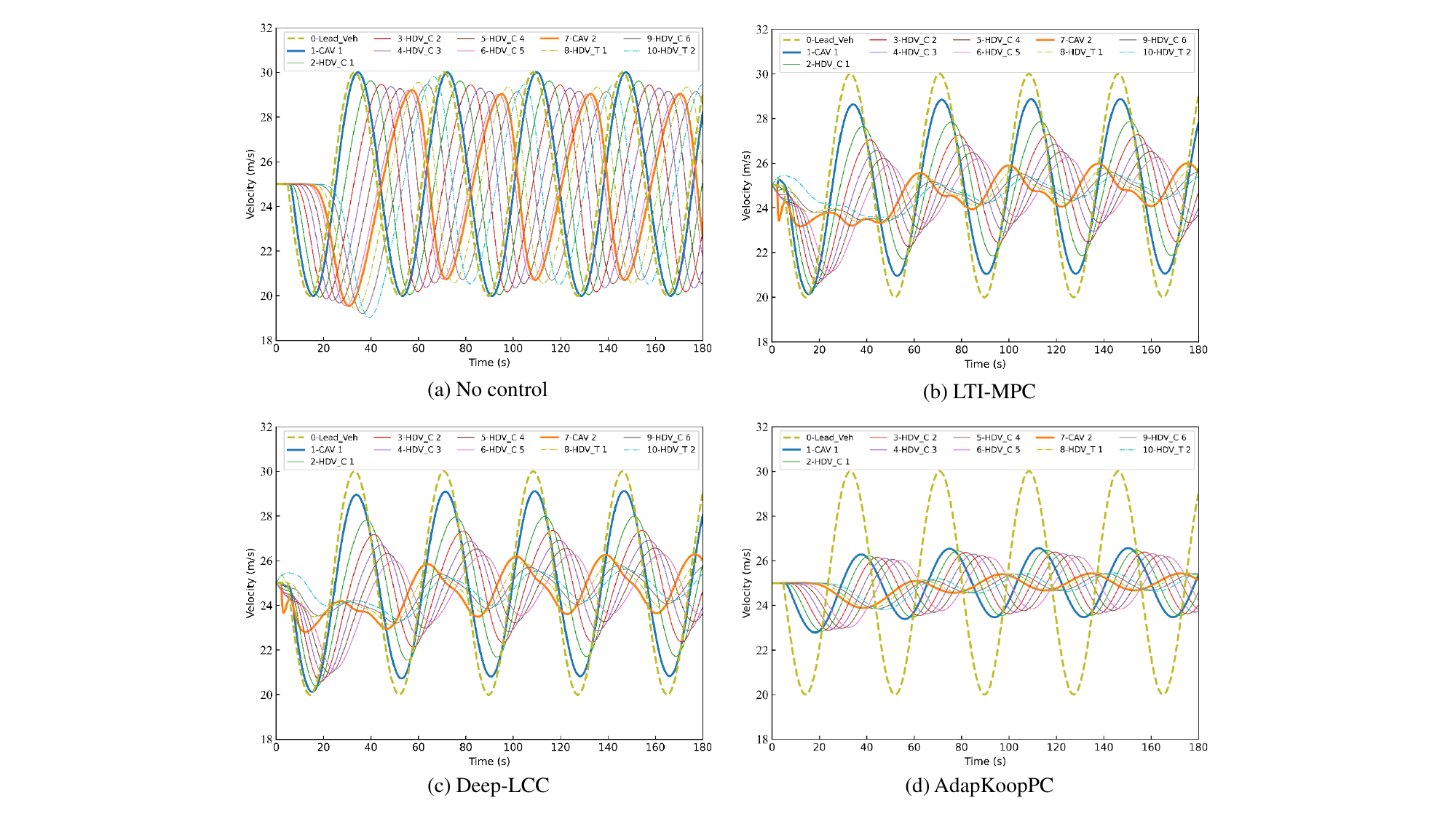}
    \vspace{-1em}
    \caption{\centering{The mixed traffic flow velocity evolution comparison under different control methods}}
	\label{Figure 9}
    \vspace{-1em}  
\end{figure}

\begin{figure}[pos=htbp]
    \vspace{1em}  
    \centering
    \includegraphics[width=0.9\textwidth]{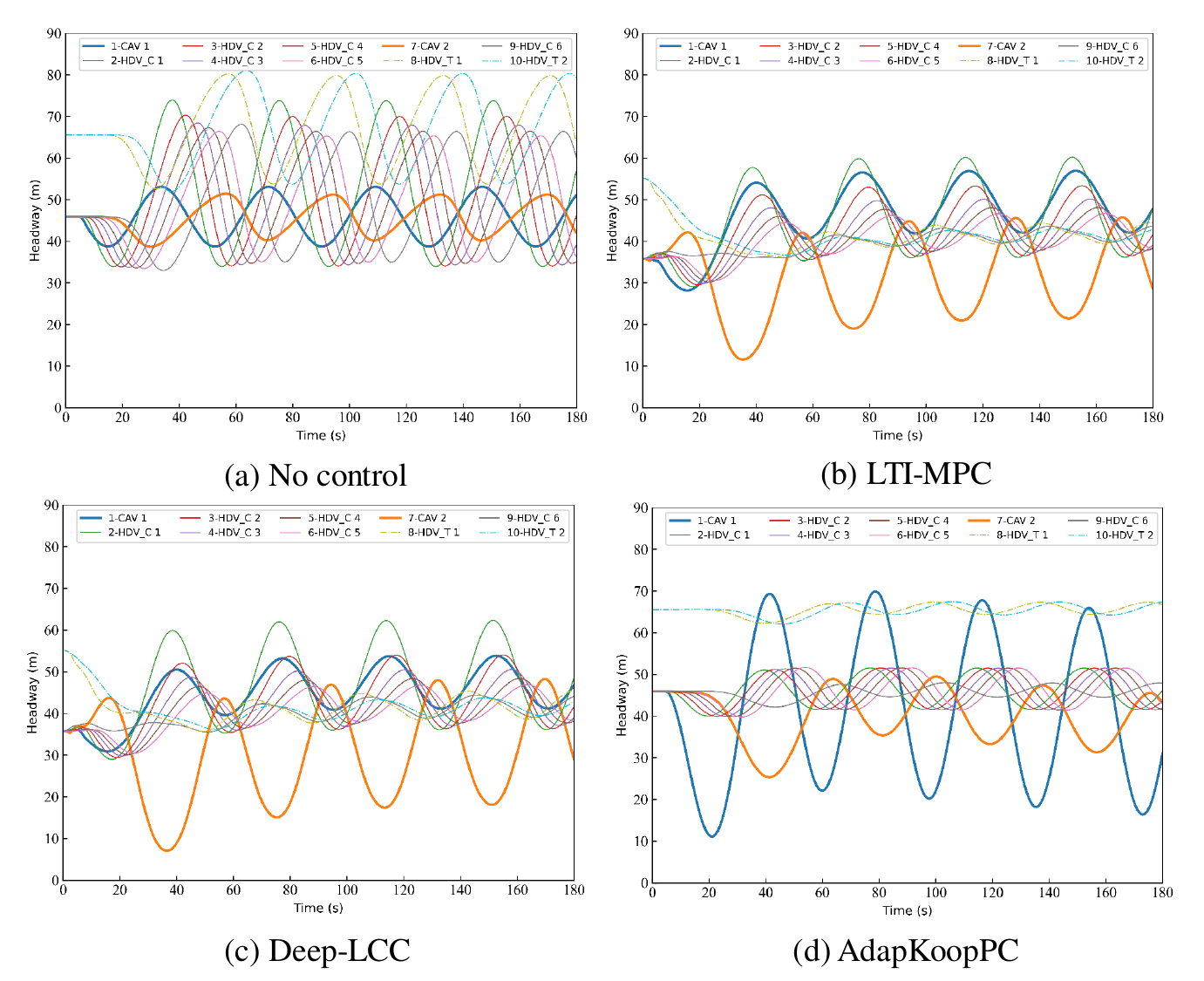}
    \vspace{-1em}
    \caption{\centering{The mixed traffic flow headway evolution comparison under different control methods}}
	\label{Figure 10}
    \vspace{-1em}  
\end{figure}

Fig. \ref{Figure 9} and Fig. \ref{Figure 10} depict and compares the headway and acceleration evolution of small-scale mixed traffic system under different control methods. The yellow dashed line represents the head vehicle, while the blue and orange thick solid lines represent the two CAVs under various control methods. The remaining 8 vehicles are all HDVs, with HDV8 and HDV10 being trucks and the rest being cars.
Fig. \ref{Figure 9} (a) and Fig. \ref{Figure 10} (a) depict the evolution of the mixed traffic system corresponding to no control, where the CAV solely considers optimizing its following trajectory with respect to the head vehicle, without taking into account its own guiding influence on the subsequent HDVs. After the head vehicle experiences disturbances starting at 4.8s and begins to oscillate, and the entire system exhibits substantial acceleration and deceleration. The evolution of the blue line in the middle subgraph indicates that as the CAV decelerates with the leading vehicle, the velocity of the entire mixed traffic system also begins to oscillate significantly. The following HDVs exhibit significant fluctuations in their headways. Taking the first HDV following behind as an example, the headway drops to 33 meters and then continues to rise to around 72 meters, indicating poor stability due to the promotion of disturbances through the entire system.

Fig. \ref{Figure 9} (b) and Fig. \ref{Figure 10} (b) shows the mixed traffic system evolution after implementing the LTI-MPC. There is a change that needs to be explained in advance, which is that the first CAV did not maintain a constant velocity with the leading vehicle at the beginning. This is because LTI-MPC imposes a penalty on the error between the actual headway and the expected headway. Once there is a small error between the initial headway set in the experiment and the expected headway after traffic system operates, it will lead to fluctuations in the initial stage of mixed traffic system evolution, which is difficult to be avoided in the experimental setup. It is obvious that the amplitude of the velocity oscillation of the first CAV also begins to oscillate after receiving the oscillation propagation from the leading vehicle, and the following HDV also oscillates with it. The overall oscillation amplitude decreases relative to Basic-MPC. In terms of headway, taking the second HDV indicated by the red line as an example, the oscillation amplitude of the headway improved due to the influence of the CAV ahead, reaching a maximum front distance of about 50 meters at 40 seconds. Subsequently, the second CAV also optimized its driving behavior under the action of LTI-MPC, and the trajectories of the following two trucks also fluctuated accordingly, maintaining a distance near the expected headway. 

The evolution and improvement of the mixed traffic system under the Deep-LCC control scenario are shown in Fig. \ref{Figure 9} (c) and Fig. \ref{Figure 10} (c). Similarly, the initial evolution stage, like the LTI-MPC, requires the first CAV to accelerate in order to achieve equilibrium headway. We can clearly observe that there is little difference between DeeP-LCC and LTI-MPC in overall traffic system evolution improvement, achieving performance comparable to LTI-MPC. The specific details will not be repeated in the description. Of course, Deep-LCC does not require prior knowledge of system dynamics, and its advantage lies in its data-driven non parametric strategy.

To shift the perspective shifts to Fig. \ref{Figure 9} (d) and Fig. \ref{Figure 10} (d)., the figures show how the mixed vehicle latoon trajectory evolves under AdapKoopPC control. In terms of acceleration, the CAV perceives the oscillation of the leading vehicle and begins to slow down slightly, then maintains a small amplitude of acceleration and deceleration driving behavior, and induces the vehicles behind to maintain a small amplitude of acceleration and deceleration driving behavior. The acceleration evolution is noticeably improved compared to the previous scenario, with the fluctuation range around -0.25m/s to 0.25 m/s. In terms of headway, the headway of CAV is relatively large compared to other previous control methods, but it ensures that the oscillation of the rear HDV headway is significantly reduced, from the previous 30-60 meters to 40-50 meters. This is direct evidence that AdapKoopPC effectively proactively guides and optimizes subsequent HDV driving behavior. It is worth mentioning that the headways of the two trucks after the second CAV fluctuates smoothly and is almost unaffected by the disturbance of the leading vehicle.

Overall, these comparative results further demonstrate that AdapKoopPC can effectively and proactively guide the subsequent HDVs and maintains the smooth driving of the mixed traffic system. More importantly, whether it is LTI-MPC or Deep-LCC, it is necessary to utilize a given model or simulation to generate some trajectories. However, these trajectories are highly dependent on the expected headway, which has limitations. AdapKoopPC adopts a penalty of velocity difference, which matches the expected goal of most drivers of vehicles in a following state. This will not cause discomfort to the driver and is more conducive to optimize the control of the mixed traffic flow system.

\subsubsection{Computation time comparison}
Table \ref{Tabel:4} states the real-time computing time for different control methods at each time step, with a time step of 0.12 seconds. The average computation time for LTI-MPC, Deep LCC, and AdadKoopPC is 0.004s, 0.326s, and 0.009s, respectively. To investigate its reasons, LTI-MPC does not require neural network encoding, and the mixed traffic system state dimension is low, so the solution is the fastest. As for Deep-LCC, it was shown to be 0.05s in theprevious work. The online computation time of Deep-LCC has shown significant fluctuations in our experiments. When simulating mixed traffic flow system scenarios similar to pre-generated trajectory scenarios, the average computation delay is 0.326 seconds. When there is a significant difference between the two scenarios, the model training solution needs to iterate to the maximum number of manually set iterations. Therefore, the delay depends on this value and will not be lower than 0.326 seconds, which may be due to a lack of generalization ability. After the impact of neural network encoding time and significant increase in state dimension, the online computation of AdadKoopPC yields an average time of 0.009 seconds, and the sampling interval is 0.12 seconds, which demonstrates the feasibility of implementing real-time mixed traffic system control. It is worth mentioning that the simulation experiments of LTI-MPC and Deep-LCC are based on the MATLAB quadprog optimizer, which has better solving performance compared to the SciPy adopted in AdadKoopPC. Therefore, the online computation time of AdaptKoopPC can still be further improved. 

\begin{table}[pos=h]
  \caption{Computation Time}\label{Tabel:4}
  \centering
\setlength{\tabcolsep}{10mm}{
\begin{tabular}{|c|c|c|c|}
\hline
 Model   & LTI-MPC   & Deep-LCC   & AdaptKoopPC     \\
\hline
Computing Time (s)   & 0.004  & 0.326& 0.009    \\
\hline
\end{tabular}
}
\end{table}

\subsection{Real-time control optimization experiment for large-scale mixed traffic system}
This section verifies the effectiveness and strong generalization ability of AdadKoopPC for large-scale mixed traffic system control, where the leading vehicle is followed by 50 vehicles. Three types of traffic flow scenarios covering different CAVs penetration rates, communication range degradation, and different CAVs distribution are set up to analyze the effectiveness and performance loss of AdadKoopPC in traffic flow optimization control in a wide range of scenarios.

\subsubsection{CAVs penetration rates}
\begin{figure}[pos=htbp]
    \vspace{1em}  
    \centering
    \includegraphics[width=1\textwidth]{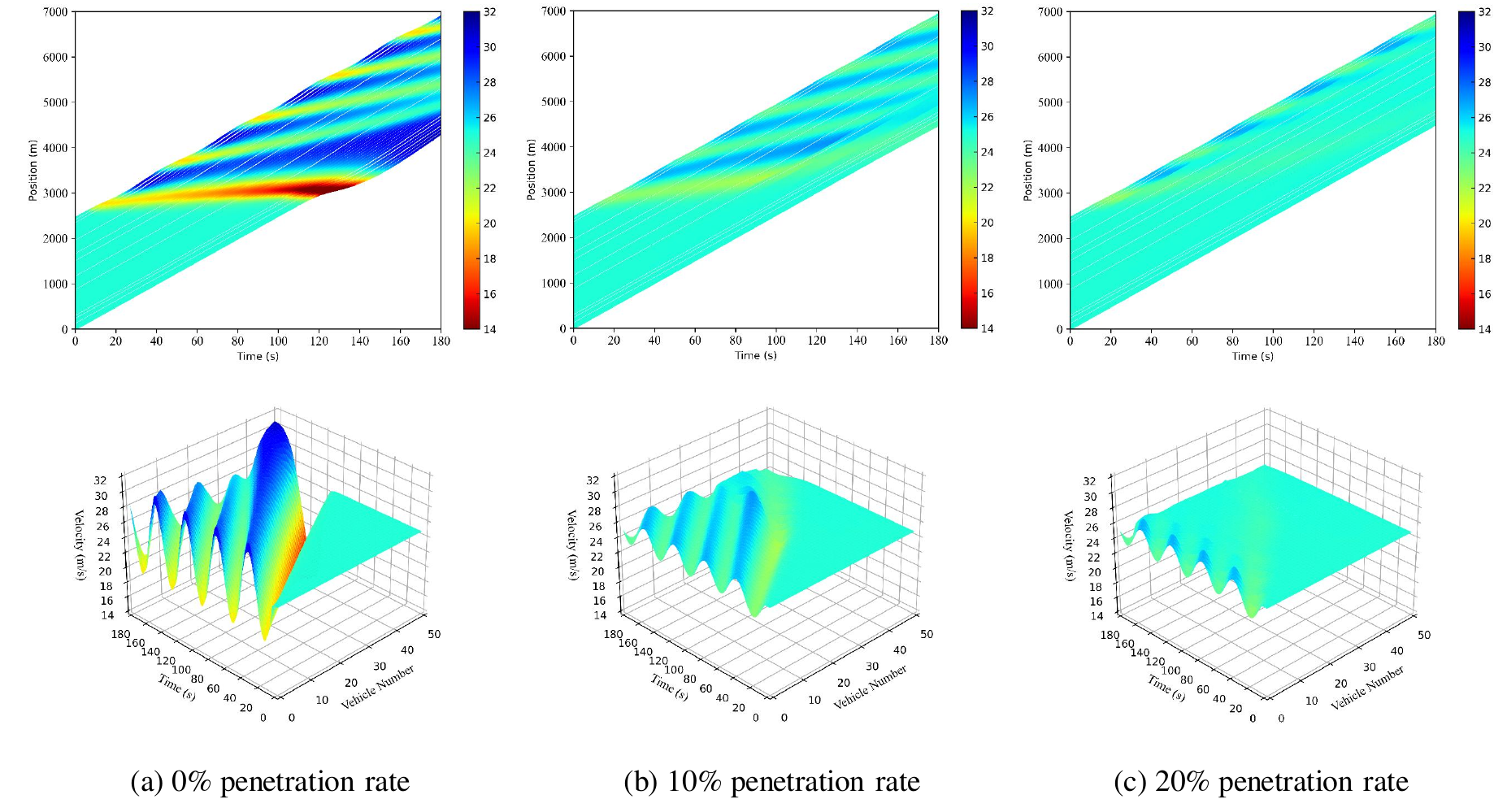}
    \vspace{-1em}
    \caption{\centering{The mixed traffic flow evolution with AdapKoopPC under different CAVs penetration rates}}
	\label{Figure 11}
    \vspace{-1em}  
\end{figure}

Fig. \ref{Figure 11}  depicts the overall evolution of the mixed vehicle with AdadKoopPC under different CAV penetration rates in the large-scale mixed traffic system scenario, including 0\%, 10\%, and 20\%. The horizontal axis represents time, the vertical axis represents vehicle position, and the color axis represents velocity. Note that there are some white spacing lines, indicating that the trajectory here is a truck trajectory with a greater headway than the headway of the car. The following row shows the corresponding three-dimensional evolution diagram. The experimental comparison results can be mainly summarized into the following two points:
a) When there is no CAV in the traffic system, the disturbance of the leading vehicle propagates directly upstream with the traffic wave, which can be reflected by the increasingly prominent red area in Fig. \ref{Figure 11}(a). The three-dimensional velocity evolution diagram provides a more intuitive view of the poor driving conditions of the traffic system.
b) As the penetration rate increases to 10\%, the traffic wave gradually dissipates and there is no significant fluctuation in the velocity of following vehicles. Until the penetration rate reaches 20\%, the velocity oscillation can be almost ignored and maintained around the expected velocity of 25m/s. An experiment with the 30\% penetration rate has also been conducted, and its effect is very close to that of 20\%, which is not shown here. It can be considered that the effect is approaching saturation at 20\% penetration rate. Overall, AdadKoopPC has the ability to significantly alleviate upstream disturbance waves.

\subsubsection{Communication range degradation}
\begin{figure}[pos=htbp]
    \vspace{1em}  
    \centering
    \includegraphics[width=1\textwidth]{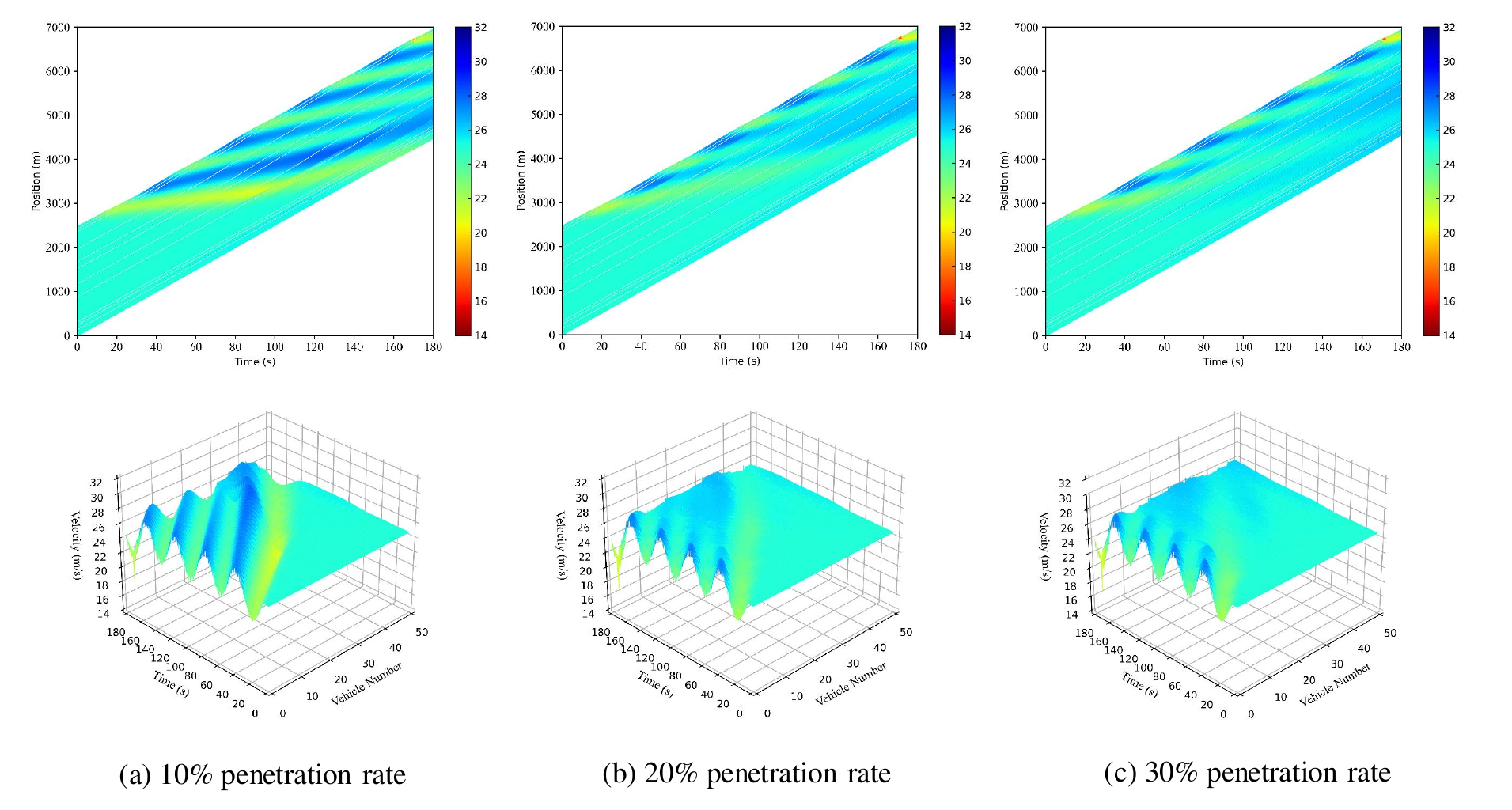}
    \vspace{-1em}
    \caption{\centering{The mixed traffic flow evolution with AdapKoopPC under communication range degradation}}
	\label{Figure 12}
    \vspace{-1em}  
\end{figure}

Another special case has also been considered, where CAV deteriorates within communication range due to external factors. The vehicle can only receive information from the following vehicle, and cannot obtain the driving states of all vehicles. And the performance degradation of AdadKoopPC in this scenario is investigated ,the results are shown in Fig. \ref{Figure 12}. Note that more communication restricted scenarios fall between the communication degradation scenario discussed here and the communication lossless scenario, so the control effect of AdadKoopPC in other scenarios will be better than the extreme scenario presented below.

 Three penetration rates of 10\%, 20\%, and 30\%  are set for experiments under the premise of communication range degradation. Firstly, by comparing Fig. \ref{Figure 12} (a) with Fig. \ref{Figure 11} (a), AdadKoopPC still plays a significant role, and the traffic oscillation wave generated by the leading vehicle is significantly alleviated. Secondly, the difference between Fig. \ref{Figure 12} (a) and Fig. \ref{Figure 11} (b) intuitively reflects the impact of communication range degradation on the overall evolution improvement of the traffic system. The oscillation slightly increases, indicating that the performance of AdadKoopPC has been slightly degraded. Similarly, in the scenario of a 20\% penetration rate, due to the impact of communication range degradation on AdadKoopPC, CAV only considers optimizing and inducing the first vehicle behind it, resulting in a decrease in the overall mixed traffic system evolution effect, which can be observed by comparing Fig. \ref{Figure 12} (b) and Fig. \ref{Figure 11} (c).
Of course, under the above two penetration rates, even if the communication range is degraded, the traffic system oscillation propagation is still within an acceptable range, proving that AdadKoopPC can still induce and optimize the performance of the entire system even under the influence of communication range degradation. Fig. \ref{Figure 12} (c) corresponds to a 30\% penetration rate. The simulation results confirm that the evolution of the traffic system is stable and smooth, and the optimization effect is approaching saturation. 

\subsubsection{CAVs distribution}
\begin{figure}[pos=htbp]
    \vspace{1em}  
    \centering
    \includegraphics[width=0.8\textwidth]{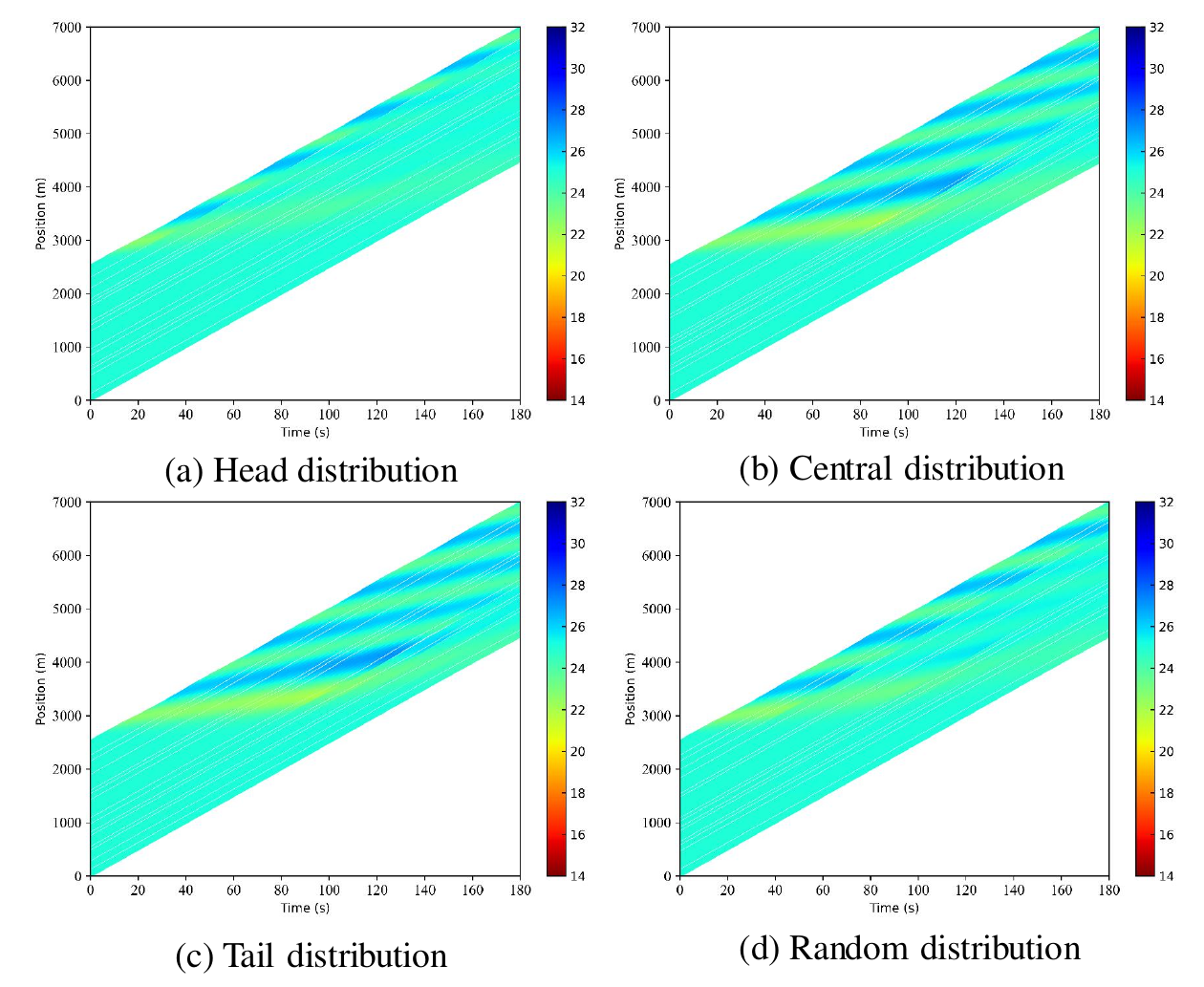}
    \vspace{-1em}
    \caption{\centering{The mixed traffic flow evolution with AdapKoopPC under different CAVs distribution}}
	\label{Figure 13}
    \vspace{-1em}  
\end{figure}

To compare the control performance of AdapKoopPC in mitigating traffic oscillations under different CAV distributions in the mixed traffic flow, the arrangements of HDVs and CAVs are randomly generated while maintaining a fixed CAV penetration rate of 10\%. Fig. \ref{Figure 13} aims to explore the impact of AdapKoopPC on alleviating traffic oscillations under different CAV distributions. Fig. \ref{Figure 13} (a) corresponds to a scenario where three CAVs are concentrated in the front of the mixed traffic flow, and its optimization effect on the mixed traffic flow is very close to Fig. \ref{Figure 11} (c). This is related to the fact that CAVs in the front of the mixed traffic flow can eliminate oscillations almost without affecting subsequent vehicles. Fig. \ref{Figure 13} (b) and Fig. \ref{Figure 13} (c) correspond to the concentration of CAVs in the middle and rear of all vehicles, respectively. It can be observed that the effect is not as good as when CAVs are located in the front, which is understandable. Yet, the overall disturbance mitigation effect is significant. Additionally, an interesting finding is that when CAVs are more evenly distributed, the mitigation performance is better than when they are concentrated in the middle or rear. Overall, regardless of the position of the CAVs, AdapKoopPC performs well, fully demonstrating its effectiveness and universality, strong generalization ability, and suitability for various scenarios.

\subsubsection{Number of controller deployments}
We further investigate how many CAVs are required to deploy the proposed AdapKoopPC controller in random conditions to achieve a satisfactory level of traffic flow improvement. A mixed traffic flow consisting of 20 CAVs and 30 HDVs were randomly generated, where the AdapKoopPC was randomly deployed on the CAVs. In order to quantitatively analyze the traffic flow improvement, the velocity standard deviation and headway standard deviation of all vehicles in the mixed traffic flow are used as evaluation indicators. The results are shown in Table  \ref{Tabel:5}. Overall, as the number of deployed controllers increases, the standard deviation of speed and headway gradually decreases. More importantly, the standard deviation of speed and headway archives convergence when the controller was 15 in such a mixed traffic flow. At such situation, the speed oscillation was reduced by 52.12\%, and the standard deviation of the headway was reduced by 53.36\%.

\begin{table}
 \vspace{-1em}
  \caption{The impact of the number of controllers on reducing traffic flow oscillations}\label{Tabel:5}
  \centering
\setlength{\tabcolsep}{1mm}{
\begin{tabular}{|c|c|c|c|c|c|}
\hline
Number  & ${v_{std}}$ (m/s) & ${h_{std}}$ (m) &  Number & ${v_{std}}$ & ${h_{std}}$  (m)  \\
\hline
0   & 3.30  & 8.34 & 1   &2.28    &5.55 \\
2   & 2.18  & 5.47 & 3   & 2.16  & 5.65\\
4   &2.07   & 5.48 & 5   & 2.04  & 5.55 \\
6   & 1.85  & 4.85  & 7   &1.84   & 4.86  \\
8   & 1.65   & 4.15 & 9   &1.61   & 3.99 \\
10  & 1.61 & 3.98 & 11   & 1.59 & 3.92  \\
12   & 1.60   & 3.94 & 13   & 1.58   &3.89 \\
14   & 1.58  & 3.88 & 15   &1.58    &3.89 \\
16   & 1.59  & 3.86 & 17   &1.59    &3.84 \\
18   & 1.59  & 3.83& 19   &1.59    &3.83 \\
20   & 1.59  & 3.81 & /   &/    &/ \\
\hline
\end{tabular}
}
 \vspace{-2em}
\end{table}

\section{Conclusions}
This study addressed the fundamental challenge of mitigating traffic oscillations in mixed traffic flow: the inherent conflict between the need for high-fidelity, nonlinear models of HDV behavior and the requirement for computationally tractable, linear models for real-time MPC. To resolve this, we proposed a novel framework based on Koopman operator theory, which successfully decouples modeling complexity from control complexity. Our proposed deep Koopman network, AdapKoopnet, demonstrated its capability to learn complex, personalized driving behaviors from naturalistic data, autonomously identifying latent traffic scenarios and driver characteristics without pre-defined labels. Its superior multi-step prediction performance, which surpasses that of traditional nonlinear deep learning models, provided the foundation for our control strategy.

The resulting \textbf{AdapKoopPC} framework integrates these learned, high-dimensional linear models into an MPC scheme, creating a scalable and adaptive control architecture for CAVs. Extensive simulations validated the efficacy of this approach. In both small- and large-scale mixed traffic systems, AdapKoopPC significantly attenuated traffic oscillations and enhanced traffic stability, proving effective even at low CAV penetration rates of 10-20\%. The framework demonstrated robustness across various CAV distributions and under degraded communication conditions, all while maintaining remarkable computational efficiency suitable for real-time deployment. This work confirms that the proposed Koopman-based data-driven predictive control paradigm offers a powerful and practical solution for smoothing mixed traffic flow.

Despite these promising results, this study has limitations that open avenues for future research. The current AdapKoopnet relies on offline training and does not support online updates, which could be critical for adapting to non-stationary driving behaviors. Moreover, the scalability of the centralized MPC in AdapKoopPC may still face challenges in ultra-large-scale systems. Future work will focus on developing online learning mechanisms for the Koopman operator and exploring efficient distributed control architectures to further enhance the framework's applicability to complex, real-world traffic environments, including multi-lane scenarios.

\section*{Acknowledgement}
This research was supported by the project of the National Key R\&D Program of China (No. 2023YFB4302701), the National Natural Science Foundation of China (No. 52232012, 52272343, 52131203), SEU Innovation Capability Enhancement Plan for Doctoral Students (No.CXJH\_SEU 25178) and Postgraduate Research \& Practice Innovation Program of Jiangsu Province (No. KYCX24\_0452).

\bibliographystyle{cas-model2-names}

\bibliography{main}

\begin{thebibliography}{71}
\expandafter\ifx\csname natexlab\endcsname\relax\def\natexlab#1{#1}\fi
\providecommand{\url}[1]{\texttt{#1}}
\providecommand{\href}[2]{#2}
\providecommand{\path}[1]{#1}
\providecommand{\DOIprefix}{doi:}
\providecommand{\ArXivprefix}{arXiv:}
\providecommand{\URLprefix}{URL: }
\providecommand{\Pubmedprefix}{pmid:}
\providecommand{\doi}[1]{\href{http://dx.doi.org/#1}{\path{#1}}}
\providecommand{\Pubmed}[1]{\href{pmid:#1}{\path{#1}}}
\providecommand{\bibinfo}[2]{#2}
\ifx\xfnm\relax \def\xfnm[#1]{\unskip,\space#1}\fi
\bibitem[{Avila and Mezi{\'c}(2020)}]{avila2020data}
\bibinfo{author}{Avila, A.M.}, \bibinfo{author}{Mezi{\'c}, I.}, \bibinfo{year}{2020}.
\newblock \bibinfo{title}{Data-driven analysis and forecasting of highway traffic dynamics}.
\newblock \bibinfo{journal}{Nature communications} \bibinfo{volume}{11}, \bibinfo{pages}{2090}.
\bibitem[{Ba et~al.(2016)Ba, Kiros and Hinton}]{ba2016layer}
\bibinfo{author}{Ba, J.L.}, \bibinfo{author}{Kiros, J.R.}, \bibinfo{author}{Hinton, G.E.}, \bibinfo{year}{2016}.
\newblock \bibinfo{title}{Layer normalization}.
\newblock \bibinfo{journal}{arXiv preprint arXiv:1607.06450} .
\bibitem[{Bando et~al.(1998)Bando, Hasebe, Nakanishi and Nakayama}]{bando1998analysis}
\bibinfo{author}{Bando, M.}, \bibinfo{author}{Hasebe, K.}, \bibinfo{author}{Nakanishi, K.}, \bibinfo{author}{Nakayama, A.}, \bibinfo{year}{1998}.
\newblock \bibinfo{title}{Analysis of optimal velocity model with explicit delay}.
\newblock \bibinfo{journal}{Physical Review E} \bibinfo{volume}{58}, \bibinfo{pages}{5429}.
\bibitem[{Chen et~al.(2024a)Chen, Wang, Hu, He, Yan, Wen and Du}]{chen2024data}
\bibinfo{author}{Chen, B.}, \bibinfo{author}{Wang, M.}, \bibinfo{author}{Hu, L.}, \bibinfo{author}{He, G.}, \bibinfo{author}{Yan, H.}, \bibinfo{author}{Wen, X.}, \bibinfo{author}{Du, R.}, \bibinfo{year}{2024}a.
\newblock \bibinfo{title}{Data-driven koopman model predictive control for hybrid energy storage system of electric vehicles under vehicle-following scenarios}.
\newblock \bibinfo{journal}{Applied Energy} \bibinfo{volume}{365}, \bibinfo{pages}{123218}.
\bibitem[{Chen et~al.(2014)Chen, Ahn, Laval and Zheng}]{chen2014periodicity}
\bibinfo{author}{Chen, D.}, \bibinfo{author}{Ahn, S.}, \bibinfo{author}{Laval, J.}, \bibinfo{author}{Zheng, Z.}, \bibinfo{year}{2014}.
\newblock \bibinfo{title}{On the periodicity of traffic oscillations and capacity drop: The role of driver characteristics}.
\newblock \bibinfo{journal}{Transportation research part B: methodological} \bibinfo{volume}{59}, \bibinfo{pages}{117--136}.
\bibitem[{Chen et~al.(2024b)Chen, Yuan, Zhu, Zheng, Shen, Wang, Wang and Wang}]{chen2024aggfollower}
\bibinfo{author}{Chen, X.}, \bibinfo{author}{Yuan, X.}, \bibinfo{author}{Zhu, M.}, \bibinfo{author}{Zheng, X.}, \bibinfo{author}{Shen, S.}, \bibinfo{author}{Wang, X.}, \bibinfo{author}{Wang, Y.}, \bibinfo{author}{Wang, F.Y.}, \bibinfo{year}{2024}b.
\newblock \bibinfo{title}{Aggfollower: Aggressiveness informed car-following modeling}.
\newblock \bibinfo{journal}{IEEE Transactions on Intelligent Vehicles} .
\bibitem[{Das et~al.(2023)Das, Mustavee, Agarwal and Hasan}]{das2023koopman}
\bibinfo{author}{Das, S.}, \bibinfo{author}{Mustavee, S.}, \bibinfo{author}{Agarwal, S.}, \bibinfo{author}{Hasan, S.}, \bibinfo{year}{2023}.
\newblock \bibinfo{title}{Koopman-theoretic modeling of quasiperiodically driven systems: Example of signalized traffic corridor}.
\newblock \bibinfo{journal}{IEEE Transactions on Systems, Man, and Cybernetics: Systems} \bibinfo{volume}{53}, \bibinfo{pages}{4466--4476}.
\bibitem[{Dauphin et~al.(2017)Dauphin, Fan, Auli and Grangier}]{dauphin2017language}
\bibinfo{author}{Dauphin, Y.N.}, \bibinfo{author}{Fan, A.}, \bibinfo{author}{Auli, M.}, \bibinfo{author}{Grangier, D.}, \bibinfo{year}{2017}.
\newblock \bibinfo{title}{Language modeling with gated convolutional networks}, in: \bibinfo{booktitle}{International conference on machine learning}, \bibinfo{organization}{PMLR}. pp. \bibinfo{pages}{933--941}.
\bibitem[{Fine(2006)}]{fine2006feedforward}
\bibinfo{author}{Fine, T.L.}, \bibinfo{year}{2006}.
\newblock \bibinfo{title}{Feedforward neural network methodology}.
\newblock \bibinfo{publisher}{Springer Science \& Business Media}.
\bibitem[{Gan et~al.(2024)Gan, Chu, Li, Tang and Li}]{gan2024large}
\bibinfo{author}{Gan, L.}, \bibinfo{author}{Chu, W.}, \bibinfo{author}{Li, G.}, \bibinfo{author}{Tang, X.}, \bibinfo{author}{Li, K.}, \bibinfo{year}{2024}.
\newblock \bibinfo{title}{Large models for intelligent transportation systems and autonomous vehicles: A survey}.
\newblock \bibinfo{journal}{Advanced Engineering Informatics} \bibinfo{volume}{62}, \bibinfo{pages}{102786}.
\bibitem[{Gu et~al.(2023)Gu, Zhou and Wu}]{gu2023deep}
\bibinfo{author}{Gu, C.}, \bibinfo{author}{Zhou, T.}, \bibinfo{author}{Wu, C.}, \bibinfo{year}{2023}.
\newblock \bibinfo{title}{Deep koopman traffic modeling for freeway ramp metering}.
\newblock \bibinfo{journal}{IEEE Transactions on Intelligent Transportation Systems} .
\bibitem[{Hart et~al.(2024)Hart, Okhrin and Treiber}]{hart2024towards}
\bibinfo{author}{Hart, F.}, \bibinfo{author}{Okhrin, O.}, \bibinfo{author}{Treiber, M.}, \bibinfo{year}{2024}.
\newblock \bibinfo{title}{Towards robust car-following based on deep reinforcement learning}.
\newblock \bibinfo{journal}{Transportation research part C: emerging technologies} \bibinfo{volume}{159}, \bibinfo{pages}{104486}.
\bibitem[{He et~al.(2016)He, Zheng, Song and Zhu}]{he2016jam}
\bibinfo{author}{He, Z.}, \bibinfo{author}{Zheng, L.}, \bibinfo{author}{Song, L.}, \bibinfo{author}{Zhu, N.}, \bibinfo{year}{2016}.
\newblock \bibinfo{title}{A jam-absorption driving strategy for mitigating traffic oscillations}.
\newblock \bibinfo{journal}{IEEE Transactions on Intelligent Transportation Systems} \bibinfo{volume}{18}, \bibinfo{pages}{802--813}.
\bibitem[{Jiao et~al.(2024)Jiao, Zhai, Peng, Liu, Liang and Yin}]{jiao2024digital}
\bibinfo{author}{Jiao, Y.}, \bibinfo{author}{Zhai, X.}, \bibinfo{author}{Peng, L.}, \bibinfo{author}{Liu, J.}, \bibinfo{author}{Liang, Y.}, \bibinfo{author}{Yin, Z.}, \bibinfo{year}{2024}.
\newblock \bibinfo{title}{A digital twin-based motion forecasting framework for preemptive risk monitoring}.
\newblock \bibinfo{journal}{Advanced Engineering Informatics} \bibinfo{volume}{59}, \bibinfo{pages}{102250}.
\bibitem[{Kang et~al.(2023)Kang, Kim, Jeong and Sohn}]{kang2023trajectory}
\bibinfo{author}{Kang, Y.}, \bibinfo{author}{Kim, G.}, \bibinfo{author}{Jeong, S.}, \bibinfo{author}{Sohn, K.}, \bibinfo{year}{2023}.
\newblock \bibinfo{title}{Trajectory-based embedding for random coefficients of a theory-based car-following model}.
\newblock \bibinfo{journal}{Transportation research part C: emerging technologies} \bibinfo{volume}{152}, \bibinfo{pages}{104183}.
\bibitem[{Kim and Yeo(2024)}]{kim2024asymmetric}
\bibinfo{author}{Kim, Y.}, \bibinfo{author}{Yeo, H.}, \bibinfo{year}{2024}.
\newblock \bibinfo{title}{Asymmetric repulsive force model: a new car-following model with psycho-physical characteristics}.
\newblock \bibinfo{journal}{Transportation research part C: emerging technologies} \bibinfo{volume}{161}, \bibinfo{pages}{104571}.
\bibitem[{Koopman(1931)}]{koopman1931hamiltonian}
\bibinfo{author}{Koopman, B.O.}, \bibinfo{year}{1931}.
\newblock \bibinfo{title}{Hamiltonian systems and transformation in hilbert space}.
\newblock \bibinfo{journal}{Proceedings of the National Academy of Sciences} \bibinfo{volume}{17}, \bibinfo{pages}{315--318}.
\bibitem[{Lai(2024)}]{lai2024advancements}
\bibinfo{author}{Lai, H.Y.}, \bibinfo{year}{2024}.
\newblock \bibinfo{title}{Advancements in intelligent driving assistance: A machine learning approach to identify real-time driving strategies using environmental, eye movement, control-related, and kinetic-related data}.
\newblock \bibinfo{journal}{Advanced Engineering Informatics} \bibinfo{volume}{62}, \bibinfo{pages}{102745}.
\bibitem[{Li et~al.(2021)Li, Chen, Cao, Qu, Cheng and Li}]{li2021extraction}
\bibinfo{author}{Li, G.}, \bibinfo{author}{Chen, Y.}, \bibinfo{author}{Cao, D.}, \bibinfo{author}{Qu, X.}, \bibinfo{author}{Cheng, B.}, \bibinfo{author}{Li, K.}, \bibinfo{year}{2021}.
\newblock \bibinfo{title}{Extraction of descriptive driving patterns from driving data using unsupervised algorithms}.
\newblock \bibinfo{journal}{Mechanical Systems and Signal Processing} \bibinfo{volume}{156}, \bibinfo{pages}{107589}.
\bibitem[{Li et~al.(2024)Li, Cao and Li}]{li2024augmented}
\bibinfo{author}{Li, M.}, \bibinfo{author}{Cao, Z.}, \bibinfo{author}{Li, Z.}, \bibinfo{year}{2024}.
\newblock \bibinfo{title}{Augmented mixed vehicular platoon control with dense communication reinforcement learning for traffic oscillation alleviation}.
\newblock \bibinfo{journal}{IEEE Internet of Things Journal} \bibinfo{volume}{11}, \bibinfo{pages}{35989--36001}.
\bibitem[{Li et~al.(2025)Li, Zhou, Wang, Yang, Xu, Wang and Li}]{li2025robust}
\bibinfo{author}{Li, S.}, \bibinfo{author}{Zhou, J.}, \bibinfo{author}{Wang, J.}, \bibinfo{author}{Yang, K.}, \bibinfo{author}{Xu, Q.}, \bibinfo{author}{Wang, J.}, \bibinfo{author}{Li, K.}, \bibinfo{year}{2025}.
\newblock \bibinfo{title}{Robust explicit data-driven predictive control for mixed vehicle platoons}.
\newblock \bibinfo{journal}{IEEE Internet of Things Journal} .
\bibitem[{Li et~al.(2014)Li, Cui, An and Parsafard}]{li2014stop}
\bibinfo{author}{Li, X.}, \bibinfo{author}{Cui, J.}, \bibinfo{author}{An, S.}, \bibinfo{author}{Parsafard, M.}, \bibinfo{year}{2014}.
\newblock \bibinfo{title}{Stop-and-go traffic analysis: Theoretical properties, environmental impacts and oscillation mitigation}.
\newblock \bibinfo{journal}{Transportation Research Part B: Methodological} \bibinfo{volume}{70}, \bibinfo{pages}{319--339}.
\bibitem[{Ling et~al.(2020)Ling, Zheng, Ratliff and Coogan}]{ling2020koopman}
\bibinfo{author}{Ling, E.}, \bibinfo{author}{Zheng, L.}, \bibinfo{author}{Ratliff, L.J.}, \bibinfo{author}{Coogan, S.}, \bibinfo{year}{2020}.
\newblock \bibinfo{title}{Koopman operator applications in signalized traffic systems}.
\newblock \bibinfo{journal}{IEEE Transactions on Intelligent Transportation Systems} \bibinfo{volume}{23}, \bibinfo{pages}{3214--3225}.
\bibitem[{Liu et~al.(2025)Liu, Zheng, Liu and Liu}]{liu2025optimizing}
\bibinfo{author}{Liu, C.}, \bibinfo{author}{Zheng, F.}, \bibinfo{author}{Liu, H.X.}, \bibinfo{author}{Liu, X.}, \bibinfo{year}{2025}.
\newblock \bibinfo{title}{Optimizing mixed traffic flow: Longitudinal control of connected and automated vehicles to mitigate traffic oscillations}.
\newblock \bibinfo{journal}{IEEE Transactions on Intelligent Transportation Systems} .
\bibitem[{Liu et~al.(2019)Liu, Johns and Davison}]{liu2019end}
\bibinfo{author}{Liu, S.}, \bibinfo{author}{Johns, E.}, \bibinfo{author}{Davison, A.J.}, \bibinfo{year}{2019}.
\newblock \bibinfo{title}{End-to-end multi-task learning with attention}, in: \bibinfo{booktitle}{Proceedings of the IEEE/CVF conference on computer vision and pattern recognition}, pp. \bibinfo{pages}{1871--1880}.
\bibitem[{Long et~al.(2024)Long, Liang, Shi, Shi, Chen and Li}]{long2024traffic}
\bibinfo{author}{Long, K.}, \bibinfo{author}{Liang, Z.}, \bibinfo{author}{Shi, H.}, \bibinfo{author}{Shi, L.}, \bibinfo{author}{Chen, S.}, \bibinfo{author}{Li, X.}, \bibinfo{year}{2024}.
\newblock \bibinfo{title}{Traffic oscillation mitigation with physics-enhanced residual learning (perl)-based predictive control}.
\newblock \bibinfo{journal}{Communications in Transportation Research} \bibinfo{volume}{4}, \bibinfo{pages}{100154}.
\bibitem[{Lusch et~al.(2018)Lusch, Kutz and Brunton}]{lusch2018deep}
\bibinfo{author}{Lusch, B.}, \bibinfo{author}{Kutz, J.N.}, \bibinfo{author}{Brunton, S.L.}, \bibinfo{year}{2018}.
\newblock \bibinfo{title}{Deep learning for universal linear embeddings of nonlinear dynamics}.
\newblock \bibinfo{journal}{Nature communications} \bibinfo{volume}{9}, \bibinfo{pages}{4950}.
\bibitem[{Lyu et~al.(2025)Lyu, Guo, Liu and Wang}]{10878999}
\bibinfo{author}{Lyu, H.}, \bibinfo{author}{Guo, Y.}, \bibinfo{author}{Liu, P.}, \bibinfo{author}{Wang, T.}, \bibinfo{year}{2025}.
\newblock \bibinfo{title}{Uncertainty-aware dynamics modeling and data-driven robust predictive control for mixed vehicle platoon}.
\newblock \bibinfo{journal}{IEEE Internet of Things Journal} \bibinfo{volume}{12}, \bibinfo{pages}{17948--17963}.
\bibitem[{Lyu et~al.(2024)Lyu, Guo, Liu, Wu, Yue and Wang}]{lyu2024kooplcc}
\bibinfo{author}{Lyu, H.}, \bibinfo{author}{Guo, Y.}, \bibinfo{author}{Liu, P.}, \bibinfo{author}{Wu, Y.}, \bibinfo{author}{Yue, Q.}, \bibinfo{author}{Wang, T.}, \bibinfo{year}{2024}.
\newblock \bibinfo{title}{Kooplcc: The koopman operator-based predictive leading cruise control for mixed vehicle platoons considering the driving styles}, in: \bibinfo{booktitle}{2024 IEEE 27th International Conference on Intelligent Transportation Systems (ITSC)}, \bibinfo{organization}{IEEE}. pp. \bibinfo{pages}{01--06}.
\bibitem[{Mo et~al.(2021)Mo, Shi and Di}]{mo2021physics}
\bibinfo{author}{Mo, Z.}, \bibinfo{author}{Shi, R.}, \bibinfo{author}{Di, X.}, \bibinfo{year}{2021}.
\newblock \bibinfo{title}{A physics-informed deep learning paradigm for car-following models}.
\newblock \bibinfo{journal}{Transportation research part C: emerging technologies} \bibinfo{volume}{130}, \bibinfo{pages}{103240}.
\bibitem[{Nishi et~al.(2013)Nishi, Tomoeda, Shimura and Nishinari}]{nishi2013theory}
\bibinfo{author}{Nishi, R.}, \bibinfo{author}{Tomoeda, A.}, \bibinfo{author}{Shimura, K.}, \bibinfo{author}{Nishinari, K.}, \bibinfo{year}{2013}.
\newblock \bibinfo{title}{Theory of jam-absorption driving}.
\newblock \bibinfo{journal}{Transportation Research Part B: Methodological} \bibinfo{volume}{50}, \bibinfo{pages}{116--129}.
\bibitem[{Park et~al.(2018)Park, Kim, Kang, Chung and Choi}]{park2018sequence}
\bibinfo{author}{Park, S.H.}, \bibinfo{author}{Kim, B.}, \bibinfo{author}{Kang, C.M.}, \bibinfo{author}{Chung, C.C.}, \bibinfo{author}{Choi, J.W.}, \bibinfo{year}{2018}.
\newblock \bibinfo{title}{Sequence-to-sequence prediction of vehicle trajectory via lstm encoder-decoder architecture}, in: \bibinfo{booktitle}{2018 IEEE intelligent vehicles symposium (IV)}, \bibinfo{organization}{IEEE}. pp. \bibinfo{pages}{1672--1678}.
\bibitem[{Proctor et~al.(2018)Proctor, Brunton and Kutz}]{proctor2018generalizing}
\bibinfo{author}{Proctor, J.L.}, \bibinfo{author}{Brunton, S.L.}, \bibinfo{author}{Kutz, J.N.}, \bibinfo{year}{2018}.
\newblock \bibinfo{title}{Generalizing koopman theory to allow for inputs and control}.
\newblock \bibinfo{journal}{SIAM Journal on Applied Dynamical Systems} \bibinfo{volume}{17}, \bibinfo{pages}{909--930}.
\bibitem[{Punzo et~al.(2021)Punzo, Zheng and Montanino}]{punzo2021calibration}
\bibinfo{author}{Punzo, V.}, \bibinfo{author}{Zheng, Z.}, \bibinfo{author}{Montanino, M.}, \bibinfo{year}{2021}.
\newblock \bibinfo{title}{About calibration of car-following dynamics of automated and human-driven vehicles: Methodology, guidelines and codes}.
\newblock \bibinfo{journal}{Transportation Research Part C: Emerging Technologies} \bibinfo{volume}{128}, \bibinfo{pages}{103165}.
\bibitem[{Ruan et~al.(2022)Ruan, Wang, Zhou, Zhang, Dong and Zuo}]{ruan2022impacts}
\bibinfo{author}{Ruan, T.}, \bibinfo{author}{Wang, H.}, \bibinfo{author}{Zhou, L.}, \bibinfo{author}{Zhang, Y.}, \bibinfo{author}{Dong, C.}, \bibinfo{author}{Zuo, Z.}, \bibinfo{year}{2022}.
\newblock \bibinfo{title}{Impacts of information flow topology on traffic dynamics of cav-mv heterogeneous flow}.
\newblock \bibinfo{journal}{IEEE Transactions on Intelligent Transportation Systems} \bibinfo{volume}{23}, \bibinfo{pages}{20820--20835}.
\bibitem[{Saifuzzaman and Zheng(2014)}]{saifuzzaman2014incorporating}
\bibinfo{author}{Saifuzzaman, M.}, \bibinfo{author}{Zheng, Z.}, \bibinfo{year}{2014}.
\newblock \bibinfo{title}{Incorporating human-factors in car-following models: a review of recent developments and research needs}.
\newblock \bibinfo{journal}{Transportation research part C: emerging technologies} \bibinfo{volume}{48}, \bibinfo{pages}{379--403}.
\bibitem[{Shang et~al.(2024)Shang, Wang and Zheng}]{shang2024decentralized}
\bibinfo{author}{Shang, X.}, \bibinfo{author}{Wang, J.}, \bibinfo{author}{Zheng, Y.}, \bibinfo{year}{2024}.
\newblock \bibinfo{title}{Decentralized robust data-driven predictive control for smoothing mixed traffic flow}.
\newblock \bibinfo{journal}{IEEE Transactions on Intelligent Transportation Systems} .
\bibitem[{Shi et~al.(2025)Shi, Shi, Wu, Li, Zhou and Ran}]{shi2025predictive}
\bibinfo{author}{Shi, H.}, \bibinfo{author}{Shi, K.}, \bibinfo{author}{Wu, K.}, \bibinfo{author}{Li, W.}, \bibinfo{author}{Zhou, Y.}, \bibinfo{author}{Ran, B.}, \bibinfo{year}{2025}.
\newblock \bibinfo{title}{A predictive deep reinforcement learning based connected automated vehicle anticipatory longitudinal control in a mixed traffic lane change condition}.
\newblock \bibinfo{journal}{IEEE Internet of Things Journal} .
\bibitem[{Song et~al.(2023)Song, Zhu, Zhao, Han and Chen}]{song2023personalized}
\bibinfo{author}{Song, D.}, \bibinfo{author}{Zhu, B.}, \bibinfo{author}{Zhao, J.}, \bibinfo{author}{Han, J.}, \bibinfo{author}{Chen, Z.}, \bibinfo{year}{2023}.
\newblock \bibinfo{title}{Personalized car-following control based on a hybrid of reinforcement learning and supervised learning}.
\newblock \bibinfo{journal}{IEEE Transactions on Intelligent Transportation Systems} \bibinfo{volume}{24}, \bibinfo{pages}{6014--6029}.
\bibitem[{Stern et~al.(2018)Stern, Cui, Delle~Monache, Bhadani, Bunting, Churchill, Hamilton, Pohlmann, Wu, Piccoli et~al.}]{stern2018dissipation}
\bibinfo{author}{Stern, R.E.}, \bibinfo{author}{Cui, S.}, \bibinfo{author}{Delle~Monache, M.L.}, \bibinfo{author}{Bhadani, R.}, \bibinfo{author}{Bunting, M.}, \bibinfo{author}{Churchill, M.}, \bibinfo{author}{Hamilton, N.}, \bibinfo{author}{Pohlmann, H.}, \bibinfo{author}{Wu, F.}, \bibinfo{author}{Piccoli, B.}, et~al., \bibinfo{year}{2018}.
\newblock \bibinfo{title}{Dissipation of stop-and-go waves via control of autonomous vehicles: Field experiments}.
\newblock \bibinfo{journal}{Transportation Research Part C: Emerging Technologies} \bibinfo{volume}{89}, \bibinfo{pages}{205--221}.
\bibitem[{Tian et~al.(2024)Tian, Shi, Zhou and Li}]{tian2024physically}
\bibinfo{author}{Tian, K.}, \bibinfo{author}{Shi, H.}, \bibinfo{author}{Zhou, Y.}, \bibinfo{author}{Li, S.}, \bibinfo{year}{2024}.
\newblock \bibinfo{title}{Physically analyzable ai-based nonlinear platoon dynamics modeling during traffic oscillation: A koopman approach}.
\newblock \bibinfo{journal}{arXiv preprint arXiv:2406.14696} .
\bibitem[{Treiber et~al.(2000)Treiber, Hennecke and Helbing}]{treiber2000congested}
\bibinfo{author}{Treiber, M.}, \bibinfo{author}{Hennecke, A.}, \bibinfo{author}{Helbing, D.}, \bibinfo{year}{2000}.
\newblock \bibinfo{title}{Congested traffic states in empirical observations and microscopic simulations}.
\newblock \bibinfo{journal}{Physical review E} \bibinfo{volume}{62}, \bibinfo{pages}{1805}.
\bibitem[{Vaswani et~al.(2017)Vaswani, Shazeer, Parmar, Uszkoreit, Jones, Gomez, Kaiser and Polosukhin}]{vaswani2017attention}
\bibinfo{author}{Vaswani, A.}, \bibinfo{author}{Shazeer, N.}, \bibinfo{author}{Parmar, N.}, \bibinfo{author}{Uszkoreit, J.}, \bibinfo{author}{Jones, L.}, \bibinfo{author}{Gomez, A.N.}, \bibinfo{author}{Kaiser, {\L}.}, \bibinfo{author}{Polosukhin, I.}, \bibinfo{year}{2017}.
\newblock \bibinfo{title}{Attention is all you need}.
\newblock \bibinfo{journal}{Advances in neural information processing systems} \bibinfo{volume}{30}.
\bibitem[{Wang et~al.(2023a)Wang, Lian, Jiang, Xu, Li and Jones}]{wang2023distributed}
\bibinfo{author}{Wang, J.}, \bibinfo{author}{Lian, Y.}, \bibinfo{author}{Jiang, Y.}, \bibinfo{author}{Xu, Q.}, \bibinfo{author}{Li, K.}, \bibinfo{author}{Jones, C.N.}, \bibinfo{year}{2023}a.
\newblock \bibinfo{title}{Distributed data-driven predictive control for cooperatively smoothing mixed traffic flow}.
\newblock \bibinfo{journal}{Transportation Research Part C: Emerging Technologies} \bibinfo{volume}{155}, \bibinfo{pages}{104274}.
\bibitem[{Wang et~al.(2021)Wang, Zheng, Chen, Xu and Li}]{wang2021leading}
\bibinfo{author}{Wang, J.}, \bibinfo{author}{Zheng, Y.}, \bibinfo{author}{Chen, C.}, \bibinfo{author}{Xu, Q.}, \bibinfo{author}{Li, K.}, \bibinfo{year}{2021}.
\newblock \bibinfo{title}{Leading cruise control in mixed traffic flow: System modeling, controllability, and string stability}.
\newblock \bibinfo{journal}{IEEE Transactions on Intelligent Transportation Systems} \bibinfo{volume}{23}, \bibinfo{pages}{12861--12876}.
\bibitem[{Wang et~al.(2023b)Wang, Zheng, Li and Xu}]{wang2023deep}
\bibinfo{author}{Wang, J.}, \bibinfo{author}{Zheng, Y.}, \bibinfo{author}{Li, K.}, \bibinfo{author}{Xu, Q.}, \bibinfo{year}{2023}b.
\newblock \bibinfo{title}{Deep-lcc: Data-enabled predictive leading cruise control in mixed traffic flow}.
\newblock \bibinfo{journal}{IEEE Transactions on Control Systems Technology} .
\bibitem[{Wang et~al.(2023c)Wang, Shang, Levin and Stern}]{wang2023general}
\bibinfo{author}{Wang, S.}, \bibinfo{author}{Shang, M.}, \bibinfo{author}{Levin, M.W.}, \bibinfo{author}{Stern, R.}, \bibinfo{year}{2023}c.
\newblock \bibinfo{title}{A general approach to smoothing nonlinear mixed traffic via control of autonomous vehicles}.
\newblock \bibinfo{journal}{Transportation Research Part C: Emerging Technologies} \bibinfo{volume}{146}, \bibinfo{pages}{103967}.
\bibitem[{Wang et~al.(2024a)Wang, Ngoduy, Li, Lyu, Zou and Dantsuji}]{wang2024koopman}
\bibinfo{author}{Wang, T.}, \bibinfo{author}{Ngoduy, D.}, \bibinfo{author}{Li, Y.}, \bibinfo{author}{Lyu, H.}, \bibinfo{author}{Zou, G.}, \bibinfo{author}{Dantsuji, T.}, \bibinfo{year}{2024}a.
\newblock \bibinfo{title}{Koopman theory meets graph convolutional network: Learning the complex dynamics of non-stationary highway traffic flow for spatiotemporal prediction}.
\newblock \bibinfo{journal}{Chaos, Solitons \& Fractals} \bibinfo{volume}{187}, \bibinfo{pages}{115437}.
\bibitem[{Wang et~al.(2024b)Wang, Jiang, Wu and Yao}]{wang2024mitigating}
\bibinfo{author}{Wang, Y.}, \bibinfo{author}{Jiang, Y.}, \bibinfo{author}{Wu, Y.}, \bibinfo{author}{Yao, Z.}, \bibinfo{year}{2024}b.
\newblock \bibinfo{title}{Mitigating traffic oscillation through control of connected automated vehicles: A cellular automata simulation}.
\newblock \bibinfo{journal}{Expert Systems with Applications} \bibinfo{volume}{235}, \bibinfo{pages}{121275}.
\bibitem[{Wang et~al.(2024c)Wang, Li and Yang}]{wang2024adaptive}
\bibinfo{author}{Wang, Y.}, \bibinfo{author}{Li, H.X.}, \bibinfo{author}{Yang, H.}, \bibinfo{year}{2024}c.
\newblock \bibinfo{title}{Adaptive spatial-model-based predictive control for complex distributed parameter systems}.
\newblock \bibinfo{journal}{Advanced Engineering Informatics} \bibinfo{volume}{59}, \bibinfo{pages}{102331}.
\bibitem[{Wang et~al.(2025)Wang, Wei, Tang, Zhao, Hu and Zhang}]{wang2025adaptive}
\bibinfo{author}{Wang, Z.}, \bibinfo{author}{Wei, C.}, \bibinfo{author}{Tang, X.}, \bibinfo{author}{Zhao, W.}, \bibinfo{author}{Hu, C.}, \bibinfo{author}{Zhang, X.}, \bibinfo{year}{2025}.
\newblock \bibinfo{title}{Adaptive risk tendency in uncertainty-aware motion planning using risk-sensitive reinforcement learning}.
\newblock \bibinfo{journal}{Advanced Engineering Informatics} \bibinfo{volume}{63}, \bibinfo{pages}{102942}.
\bibitem[{Williams et~al.(2015)Williams, Kevrekidis and Rowley}]{williams2015data}
\bibinfo{author}{Williams, M.O.}, \bibinfo{author}{Kevrekidis, I.G.}, \bibinfo{author}{Rowley, C.W.}, \bibinfo{year}{2015}.
\newblock \bibinfo{title}{A data--driven approximation of the koopman operator: Extending dynamic mode decomposition}.
\newblock \bibinfo{journal}{Journal of Nonlinear Science} \bibinfo{volume}{25}, \bibinfo{pages}{1307--1346}.
\bibitem[{Xiao et~al.(2022)Xiao, Zhang, Xu, Liu and Liu}]{xiao2022deep}
\bibinfo{author}{Xiao, Y.}, \bibinfo{author}{Zhang, X.}, \bibinfo{author}{Xu, X.}, \bibinfo{author}{Liu, X.}, \bibinfo{author}{Liu, J.}, \bibinfo{year}{2022}.
\newblock \bibinfo{title}{Deep neural networks with koopman operators for modeling and control of autonomous vehicles}.
\newblock \bibinfo{journal}{IEEE Transactions on Intelligent Vehicles} \bibinfo{volume}{8}, \bibinfo{pages}{135--146}.
\bibitem[{Xu et~al.(2024a)Xu, Chen, Zhang, Wang, Liu and Guo}]{xu2024sequence}
\bibinfo{author}{Xu, N.}, \bibinfo{author}{Chen, C.}, \bibinfo{author}{Zhang, Y.}, \bibinfo{author}{Wang, J.}, \bibinfo{author}{Liu, Q.}, \bibinfo{author}{Guo, C.}, \bibinfo{year}{2024}a.
\newblock \bibinfo{title}{A sequence-to-sequence car-following model for addressing driver reaction delay and cumulative error in multi-step prediction}.
\newblock \bibinfo{journal}{IEEE transactions on intelligent transportation systems} \bibinfo{volume}{25}, \bibinfo{pages}{12203--12215}.
\bibitem[{Xu et~al.(2024b)Xu, Shi, Tong, Chen and Ge}]{xu2024multi}
\bibinfo{author}{Xu, Y.}, \bibinfo{author}{Shi, Y.}, \bibinfo{author}{Tong, X.}, \bibinfo{author}{Chen, S.}, \bibinfo{author}{Ge, Y.}, \bibinfo{year}{2024}b.
\newblock \bibinfo{title}{A multi-agent reinforcement learning based control method for cavs in a mixed platoon}.
\newblock \bibinfo{journal}{IEEE Transactions on Vehicular Technology} \bibinfo{volume}{73}, \bibinfo{pages}{16160--16172}.
\bibitem[{Zhan et~al.(2022)Zhan, Ma and Zhang}]{zhan2022data}
\bibinfo{author}{Zhan, J.}, \bibinfo{author}{Ma, Z.}, \bibinfo{author}{Zhang, L.}, \bibinfo{year}{2022}.
\newblock \bibinfo{title}{Data-driven modeling and distributed predictive control of mixed vehicle platoons}.
\newblock \bibinfo{journal}{IEEE Transactions on Intelligent Vehicles} \bibinfo{volume}{8}, \bibinfo{pages}{572--582}.
\bibitem[{Zhang et~al.(2024a)Zhang, Wang and Sun}]{zhang2024calibrating}
\bibinfo{author}{Zhang, C.}, \bibinfo{author}{Wang, W.}, \bibinfo{author}{Sun, L.}, \bibinfo{year}{2024}a.
\newblock \bibinfo{title}{Calibrating car-following models via bayesian dynamic regression}.
\newblock \bibinfo{journal}{Transportation Research Part C: Emerging Technologies} \bibinfo{volume}{168}, \bibinfo{pages}{104719}.
\bibitem[{Zhang et~al.(2024b)Zhang, Jin, McQuade, Bayen and Piccoli}]{zhang2024car}
\bibinfo{author}{Zhang, T.T.}, \bibinfo{author}{Jin, P.J.}, \bibinfo{author}{McQuade, S.T.}, \bibinfo{author}{Bayen, A.}, \bibinfo{author}{Piccoli, B.}, \bibinfo{year}{2024}b.
\newblock \bibinfo{title}{Car-following models: A multidisciplinary review}.
\newblock \bibinfo{journal}{IEEE Transactions on Intelligent Vehicles} .
\bibitem[{Zhang et~al.(2019)Zhang, Sun, Qi and Sun}]{zhang2019simultaneous}
\bibinfo{author}{Zhang, X.}, \bibinfo{author}{Sun, J.}, \bibinfo{author}{Qi, X.}, \bibinfo{author}{Sun, J.}, \bibinfo{year}{2019}.
\newblock \bibinfo{title}{Simultaneous modeling of car-following and lane-changing behaviors using deep learning}.
\newblock \bibinfo{journal}{Transportation research part C: emerging technologies} \bibinfo{volume}{104}, \bibinfo{pages}{287--304}.
\bibitem[{Zhang et~al.(2022)Zhang, Chen, Wang, Zheng and Wu}]{zhang2022generative}
\bibinfo{author}{Zhang, Y.}, \bibinfo{author}{Chen, X.}, \bibinfo{author}{Wang, J.}, \bibinfo{author}{Zheng, Z.}, \bibinfo{author}{Wu, K.}, \bibinfo{year}{2022}.
\newblock \bibinfo{title}{A generative car-following model conditioned on driving styles}.
\newblock \bibinfo{journal}{Transportation research part C: emerging technologies} \bibinfo{volume}{145}, \bibinfo{pages}{103926}.
\bibitem[{Zhang et~al.(2024c)Zhang, Wang, Sun, Wang, Wang, Li and Wang}]{zhang2024meta}
\bibinfo{author}{Zhang, Y.}, \bibinfo{author}{Wang, X.}, \bibinfo{author}{Sun, Z.}, \bibinfo{author}{Wang, P.}, \bibinfo{author}{Wang, B.}, \bibinfo{author}{Li, L.}, \bibinfo{author}{Wang, Y.}, \bibinfo{year}{2024}c.
\newblock \bibinfo{title}{Meta koopman decomposition for time series forecasting under temporal distribution shifts}.
\newblock \bibinfo{journal}{Advanced Engineering Informatics} \bibinfo{volume}{62}, \bibinfo{pages}{102840}.
\bibitem[{Zhang et~al.(2025a)Zhang, Xu, Yin, Zhou, Hu and Guo}]{zhang2025preview}
\bibinfo{author}{Zhang, Y.}, \bibinfo{author}{Xu, N.}, \bibinfo{author}{Yin, Z.}, \bibinfo{author}{Zhou, J.}, \bibinfo{author}{Hu, M.}, \bibinfo{author}{Guo, K.}, \bibinfo{year}{2025}a.
\newblock \bibinfo{title}{A preview-based path tracking control approach for model mismatch mitigation amid extreme operational conditions}.
\newblock \bibinfo{journal}{Advanced Engineering Informatics} \bibinfo{volume}{68}, \bibinfo{pages}{103710}.
\bibitem[{Zhang et~al.(2025b)Zhang, Zhong and Yu}]{zhang2025mitigating}
\bibinfo{author}{Zhang, Y.}, \bibinfo{author}{Zhong, R.}, \bibinfo{author}{Yu, H.}, \bibinfo{year}{2025}b.
\newblock \bibinfo{title}{Mitigating stop-and-go traffic congestion with operator learning}.
\newblock \bibinfo{journal}{Transportation Research Part C: Emerging Technologies} \bibinfo{volume}{170}, \bibinfo{pages}{104928}.
\bibitem[{Zhang et~al.(2025c)Zhang, Wang, Liang, Zhao and Pan}]{zhang2025mixed}
\bibinfo{author}{Zhang, Z.}, \bibinfo{author}{Wang, W.}, \bibinfo{author}{Liang, J.}, \bibinfo{author}{Zhao, W.}, \bibinfo{author}{Pan, C.}, \bibinfo{year}{2025}c.
\newblock \bibinfo{title}{Mixed platoon hierarchical control: Elevating safety, stability, and efficiency in cav-hv integration}.
\newblock \bibinfo{journal}{IEEE Transactions on Intelligent Transportation Systems} .
\bibitem[{Zhao et~al.(2024)Zhao, Meng, Li, Xu, Li and Galland}]{zhao2024survey}
\bibinfo{author}{Zhao, H.}, \bibinfo{author}{Meng, M.}, \bibinfo{author}{Li, X.}, \bibinfo{author}{Xu, J.}, \bibinfo{author}{Li, L.}, \bibinfo{author}{Galland, S.}, \bibinfo{year}{2024}.
\newblock \bibinfo{title}{A survey of autonomous driving frameworks and simulators}.
\newblock \bibinfo{journal}{Advanced Engineering Informatics} \bibinfo{volume}{62}, \bibinfo{pages}{102850}.
\bibitem[{Zheng et~al.(2024)Zheng, Zhou, Han, Li and Yu}]{zheng2024interpretable}
\bibinfo{author}{Zheng, H.}, \bibinfo{author}{Zhou, T.}, \bibinfo{author}{Han, T.}, \bibinfo{author}{Li, S.}, \bibinfo{author}{Yu, C.}, \bibinfo{year}{2024}.
\newblock \bibinfo{title}{An interpretable prediction framework for multi-class situational awareness in conditionally automated driving}.
\newblock \bibinfo{journal}{Advanced Engineering Informatics} \bibinfo{volume}{62}, \bibinfo{pages}{102683}.
\bibitem[{Zheng et~al.(2020)Zheng, Wang and Li}]{zheng2020smoothing}
\bibinfo{author}{Zheng, Y.}, \bibinfo{author}{Wang, J.}, \bibinfo{author}{Li, K.}, \bibinfo{year}{2020}.
\newblock \bibinfo{title}{Smoothing traffic flow via control of autonomous vehicles}.
\newblock \bibinfo{journal}{IEEE Internet of Things Journal} \bibinfo{volume}{7}, \bibinfo{pages}{3882--3896}.
\bibitem[{Zhou et~al.(2024)Zhou, Yan and Yang}]{zhou2024enhancing}
\bibinfo{author}{Zhou, J.}, \bibinfo{author}{Yan, L.}, \bibinfo{author}{Yang, K.}, \bibinfo{year}{2024}.
\newblock \bibinfo{title}{Enhancing system-level safety in mixed-autonomy platoon via safe reinforcement learning}.
\newblock \bibinfo{journal}{IEEE Transactions on Intelligent Vehicles} .
\bibitem[{Zhou et~al.(2025a)Zhou, Huang, Li, Li, Cao and Song}]{zhou2025knowledge}
\bibinfo{author}{Zhou, R.}, \bibinfo{author}{Huang, J.}, \bibinfo{author}{Li, M.}, \bibinfo{author}{Li, H.}, \bibinfo{author}{Cao, H.}, \bibinfo{author}{Song, X.}, \bibinfo{year}{2025}a.
\newblock \bibinfo{title}{Knowledge transfer from simple to complex: A safe and efficient reinforcement learning framework for autonomous driving decision-making}.
\newblock \bibinfo{journal}{Advanced Engineering Informatics} \bibinfo{volume}{65}, \bibinfo{pages}{103188}.
\bibitem[{Zhou et~al.(2025b)Zhou, Zheng, Tian, Jiang et~al.}]{zhou2025twenty}
\bibinfo{author}{Zhou, S.}, \bibinfo{author}{Zheng, S.}, \bibinfo{author}{Tian, J.}, \bibinfo{author}{Jiang, R.}, et~al., \bibinfo{year}{2025}b.
\newblock \bibinfo{title}{Twenty-five years of the intelligent driver model: Foundations, extensions, applications, and future directions}.
\newblock \bibinfo{journal}{arXiv preprint arXiv:2506.05909} .
\bibitem[{Zou et~al.(2025)Zou, Wu, Ding, Zhang, Zhang and Wu}]{zou2025analyzing}
\bibinfo{author}{Zou, Y.}, \bibinfo{author}{Wu, S.}, \bibinfo{author}{Ding, L.}, \bibinfo{author}{Zhang, Y.}, \bibinfo{author}{Zhang, S.}, \bibinfo{author}{Wu, L.}, \bibinfo{year}{2025}.
\newblock \bibinfo{title}{Analyzing mandatory and discretionary lane change interaction patterns using hidden markov model-based approaches}.
\newblock \bibinfo{journal}{Advanced Engineering Informatics} \bibinfo{volume}{66}, \bibinfo{pages}{103404}.

\end{thebibliography}

\end{sloppypar}
\end{document}